\documentclass{IEEEoj}

\usepackage{amsthm}
\usepackage{multicol,multirow}
\usepackage{graphicx,subcaption}
\usepackage{amssymb, amsmath}
\usepackage{xcolor}
\usepackage{cite}
\usepackage{hyperref}
\usepackage{enumerate}
\usepackage{esint}
\usepackage{balance}
\usepackage{textcomp}
\usepackage{bbold}
\usepackage{longtable}
\usepackage{algorithm}
\usepackage{algorithmic}
\usepackage{cuted}
\usepackage{booktabs}% http://ctan.org/pkg/booktabs

\graphicspath{{figures/}}
\usepackage{nomencl}
\makenomenclature

\def\BibTeX{{\rm B\kern-.05em{\sc i\kern-.025em b}\kern-.08em
    T\kern-.1667em\lower.7ex\hbox{E}\kern-.125emX}}
\AtBeginDocument{\definecolor{ojcolor}{cmyk}{0.93,0.59,0.15,0.02}}
\def\OJlogo{\vspace{-14pt}\includegraphics[height=20pt]{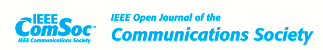}}

\begin{document}

\receiveddate{25 June, 2024}
\reviseddate{XX Month, XXXX}
\accepteddate{XX Month, XXXX}
\publisheddate{XX Month, XXXX}
\currentdate{XX Month, XXXX}
\doiinfo{OJCOMS.2022.1234567}

% !!!!!!!!!!!TBD!!!!!!!!!!!
\title{Hierarchical Blockchain Radio Access Networks: Architecture, Modelling, and Performance Assessment}

% !!!!!!!!!!!TBD!!!!!!!!!!!
\author{VASILEIOS KOUVAKIS\textsuperscript{1},
        STYLIANOS E. TREVLAKIS\textsuperscript{1},~\IEEEmembership{MEMBER,~IEEE,} 
        ALEXANDROS-APOSTOLOS A. BOULOGEORGOS\textsuperscript{2},~\IEEEmembership{SENIOR MEMBER,~IEEE,}
        HONGWU LIU\textsuperscript{3},~\IEEEmembership{SENIOR MEMBER,~IEEE,}
        WAQAS KHALID\textsuperscript{4},~\IEEEmembership{MEMBER,~IEEE,}
        THEODOROS TSIFTSIS\textsuperscript{5},~\IEEEmembership{SENIOR MEMBER,~IEEE,} and
        OCTAVIA A. DOBRE\textsuperscript{6},~\IEEEmembership{FELLOW,~IEEE}}
\affil{Department of Research and Development, InnoCube P.C., 17is Noemvriou 79, 55534 Thessaloniki, Greece}
\affil{Department of Electrical and Computer Engineering, University of Western Macedonia, ZEP Area, 50100 Kozani, Greece}
\affil{School of Information Science and Electrical Engineering, Shandong Jiaotong University, Jinan 250357, China}
\affil{Institute of Industrial Technology, Korea University, Sejong 30019, South Korea}
\affil{Department of Informatics \& Telecommunications, University of Thessaly, Lamia 35100, Greece}
\affil{Faculty of Engineering and Applied Science, Memorial University, St. John’s, NL A1B 3X9, Canada.}
\corresp{CORRESPONDING AUTHOR: Stylianos E. Trevlakis (e-mail: trevlakis@innocube.org).}
\authornote{This research was supported by the European Unions HORIZON-JU-SNS-2022 research and innovation programme under grant agreement No. 101096456, and by the Basic Science Research Program through the NRF funded by the Ministry of Education (MOE) (NRF-2022R1I1A1A01071807).}

% The paper headers
\markboth{Theoretical Performance Assessment of B-RAN}{Kouvakis \textit{et al.}}

\begin{abstract}
Demands for secure, ubiquitous, and always-available connectivity have been identified as the pillar design parameters of the next generation radio access networks (RANs).  Motivated by this, the current contribution introduces a network architecture that leverages blockchain technologies to augment security in RANs, while enabling dynamic coverage expansion through the use of intermediate commercial or private wireless nodes. To assess the efficiency and limitations of the architecture, we employ Markov chain theory in order to extract a theoretical model with increased engineering insights. Building upon this model, we quantify the latency as well as the security capabilities in terms of probability of successful attack, for three scenarios, namely fixed topology fronthaul network, advanced coverage expansion and advanced mobile node connectivity, which reveal the scalability of the blockchain-RAN architecture.

%The proposed architecture not only improves dynamic coverage, but also exhibits scalability in terms of transaction volume per block; thus, increasing its flexibility in real-world scenarios. 
%Taking advantage of Markov-chain based modeling, the performance of the system is quantified in terms of latency and security by taking into account a plethora of design parameters.
%Moreover, we apply the proposed framework in three usage scenarios, namely fronthaul network of fixed topology, advanced coverage expansion, and advanced connectivity of mobile nodes, which highlight its scalability through hierarchical deployment. Finally, numerical result verify the improved latency and flexibility without sacrificing security and resilience; therefore, demonstrating how much blockchain integration in future RAN systems increases dependability and performance.
\end{abstract}


\maketitle

\section{INTRODUCTION}
Recent technological developments and the vision for the next generation of wireless communications have brought to the forefront the need to provide intelligence, energy efficiency, and security at the far edge of the network. As a consequence, radio access networks (RANs) need to be transformed to enable flexible, efficient, and reliable connectivity to a wide variety of devices~\cite{boulogeorgos2022artificial}.
High security and privacy assurances must be implemented as RAN technologies develop to reduce newly emerging risks and vulnerabilities~\cite{Trevlakis2023}. 
This observation has motivated a great amount of research effort that aims to identify and prevent security issues in the RAN have been intensified~\cite{Rahman2022,Cao2024}. Potential threats to the confidentiality and integrity of communications include illegal access, data breaches, and network outages. 

The incorporation of blockchain to RANs presents a promising countermeasure to the above risks~\cite{Dulaimi2023}. Although blockchain had initially developed for cryptocurrency~\cite{Tschorsch2016}, its effectiveness in  decentralized tamper-proof solutions has been extensively validated~\cite{LiU2020,Wang2022,Wang2022b}. Through decentralization, the blockchain ensures that no single entity controls the network; thus, mitigating the risks associated with centralized points of failure and unauthorized access~\cite{Salman2019}.  The investigation of how blockchain can enhance security across various network areas has been widely explored making it a focal point in the evolution towards the sixth generation (6G) wireless systems. 
\color{black} Additionally, the Markovian chain technique is a versatile and powerful method that can be applied across various fields, such as marketing~\cite{Chen2021}, weather forecasting~\cite{Yutong2021}, and chemical engineering~\cite{Berthiaux2004}. An important example of its application is in blockchain technology, where it has proven effective, as demonstrated in previous studies~\cite{Kiffer2018},~\cite{Srivastava2019}. In more details, Markovian chains have been applied in blockchain within RAN (Radio Access Network) scenarios, as seen in research studies~\cite{kontorovich2022},~\cite{Li2021}. Based on these examples, we conclude that Markovian chains provide a fitting technique for this study. \color{black}
For example, the authors of~\cite{Wang2021} have conducted a comprehensive examination of the integration of blockchain into RANs, proposing a framework of secure blockchain RAN (B-RAN) tailored for 6G networking. Additionally, they  has outlined a framework for the analysis of block-structured Markov processes, adding phase-type service periods and transaction arrivals to the existing models. In~\cite{Ling2019, Ling2021}, the authors have taken advantage of Markov chain (MC) models to investigate B-RAN systems performance in terms of latency and security~capabilities. A similar modelling approach has been followed in~\cite{Sachinidis2021}, where the \color{black}authors have \color{black}presented a dual-hop B-RAN architecture, and have analyzed its performance in terms of probability of delay and average latency. 

Several recent studies have focused on the specification of the ideal block size~\cite{Li2018, Wilhelmi2022, YixinLi2020, Wilhelmi2021}. In particular, in~\cite{Li2018}, the authors have described the building and mining process with a focus on performance assessment, while the authors of~\cite{Wilhelmi2022} have studied the latency model of blockchain with a variety of timers and forks. The authors of~\cite{YixinLi2020} have presented block access control as a remedy to blockchain forking problems in wireless networks. This method effectively controls block transfer and improves transaction throughput. In the same work, an evaluation of the network performance in terms of transaction throughput and saving computational power has been conducted. Batch service queuing has been utilized in~\cite{Wilhelmi2021} in order to reduce the impact of delay on system~stability. 

\color{black}All the aforementioned works consider very constrained blockchain models that assume singular transaction per block and do not consider the possibility of block rejections. To cover this gap, this work introduce a novel B-RAN architecture and attacks modeling that allows both individual and commercial intermediary nodes to act as wireless access providers, regardless of their ownership. At the same time, the reduced complexity of the proposed model constitute it a versatile tool for conducting performance assessment and extracting design guidelines before actually deploying a B-RAN system.\color{black} In more detail, the novelty and technical contribution of the paper can be summarized~as: 
\begin{itemize}
    \item We rethink the role of connectivity providers and service consumers, and introduce a new type of B-RAN architecture, called hierarchical B-RAN (HB-RAN), which revolutionizes wireless networks, by allowing a service consumer to simultaneously be a service provider. 
    \item We identify three usage scenarios, i.e., fixed fronthaul network, advanced coverage expansion, and advanced connectivity of mobile nodes, with important business value, which are catalyzed by HB-RAN. 
    \item We provide a mathematical model that is based on MC-theory and accommodates the HB-RAN particularities. Building upon this model, we extract a closed-form expression for the average latency. 
    \item We recognize alternate history attack as the most possible type of impactful attack, and we extend the MC-based model in order to include it. Building upon  the new model, we extract the probability of successful attack, which quantifies the security capabilities of the introduced system. 
    \item Numerical results that verify the theoretical models and assessment framework as well as the engineering insights are presented. 
\end{itemize}

% The end users can access the secondary blockchain through an ad-hoc private and/or commercial network that has been created by an intermediate node. After the authorisation, all requests are sent to the primary blockchain by the intermediate node that is connected to both blockchains. Numerical results highlight that that developed B-RAN model can accommodate a higher number of degrees of freedom; thus increasing the accuracy and flexibility of the simulations. Finally, an equilibrium between latency and security is highlighted for a variety of B-RAN deployment scenarios.

\begin{figure}
    \centering
        \includegraphics[width=0.8\linewidth]{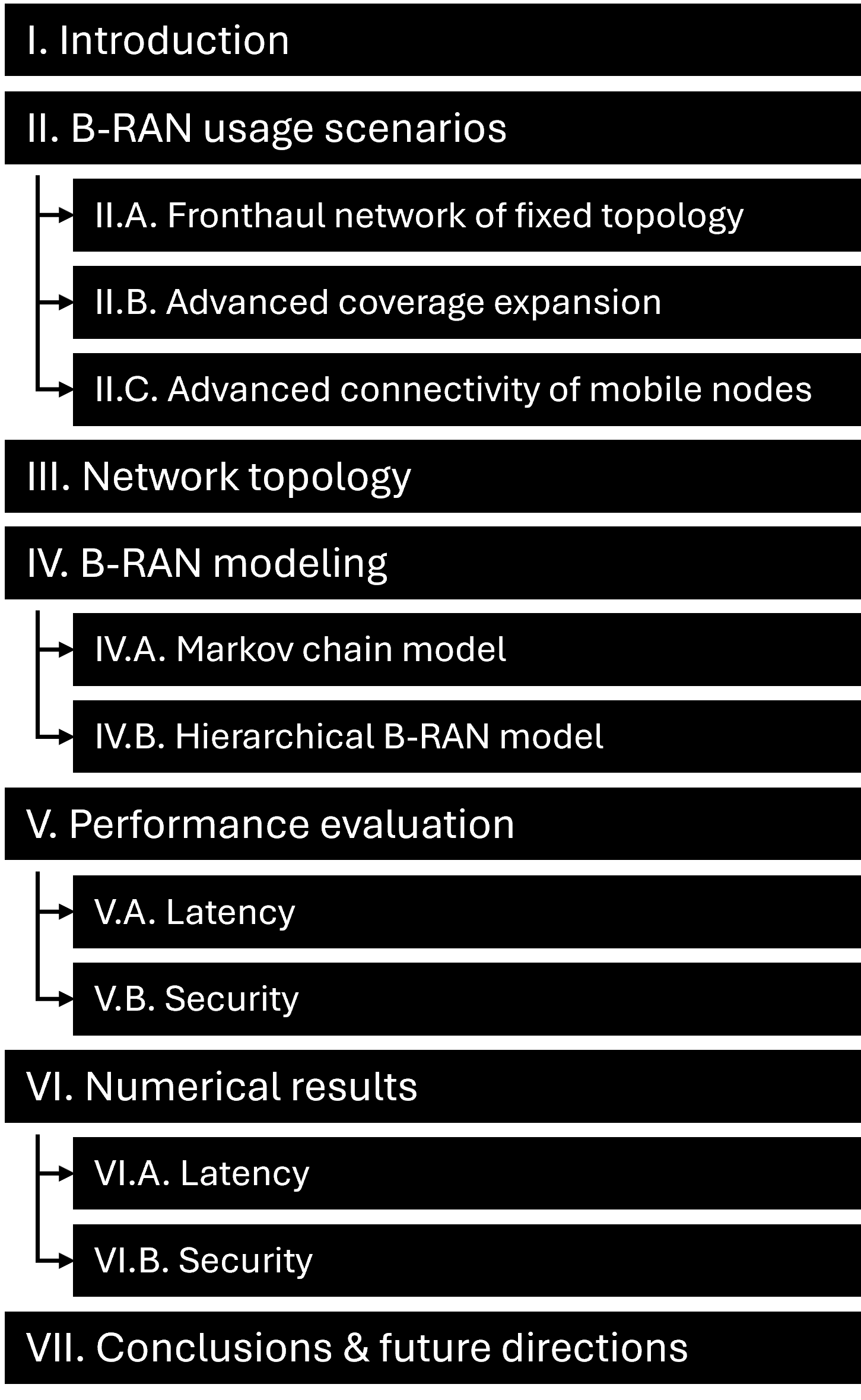}
    \caption{Paper organization.}
    \label{Fig:layout}
\end{figure}

The rest of this paper is organized as follows: The HB-RAN usage scenarios are documenteed in Section~\ref{S:bran scenarios}. A thorough system model explaining the intricate dynamics of HB-RAN is introduced in Section~\ref{S:system_model}. \color{black}In Section~\ref{S:bran_modeling}, we present the B-RAN model, employing a Markov chain model to capture the probabilistic transitions between different system states and explore their dynamic behavior during B-RAN operations. \color{black}The theoretical framework that quantifies the average latency and the probability of successful attack is provide in~Section~\ref{S:performance_evaluation}. Numerical results are presented in~Section~\ref{S:numerical_results}. Finally, concluding remarks and future research directions are provided in~Section~\ref{S:conclusions}. The organization of the paper at a glance is depicted in Fig.~\ref{Fig:layout}. 

\textit{Notations:} Unless otherwise stated, in the rest of this paper, single-server queueing models are indicated by $\text{M/M/1}$, while multi-server ones by $\text{M/M/s}$. Time is represented by $t$ and the state space of Markov chains by $E$. The probability of an event is expressed as $\Pr\{\cdot\}$. The expectation operator, which denotes the expected value of a random variable, is represented by the operator $\mathbb{E}\left[\cdot\right]$.

\makenomenclature
\nomenclature{$R_{a}$}{The rate at which an arrival event occurs.}
\nomenclature{$R_{m}$}{The rate at which a mined block event occurs.}
\nomenclature{$R_{r}$}{The rate at which request rejection events occur.}
\nomenclature{$R_{s}$}{The rate at which a service of a request occurs.}
\nomenclature{$i$}{Number of pending requests waiting to enter a block.}
\nomenclature{$j$}{Number of requests awaiting to be in service.}
\nomenclature{$k$}{Number of maximum capacity per block.}
\nomenclature{$r$}{Number of rejected requests in a block.}
\nomenclature{$h$}{Small time period that tends to infinity.}
\nomenclature{$U^{a}$}{Time between two request arrivals.}
\nomenclature{$U^{m}$}{Time between two mined events.}
\nomenclature{$U^{r}$}{Time between two rejected events.}
\nomenclature{$U^{s}$}{Time between two serviced events.}
\nomenclature{$N$}{Number of confirmations per block.}
\nomenclature{$s$}{Number of concurrent access links.}
\nomenclature{$\beta$}{The relative mining rate of the attacker.}
\nomenclature{$N_g$}{Threshold for length difference between chains.}
\nomenclature{$\rho$}{Traffic intensity.}
\nomenclature{$P_{i,j}$}{Probability to be in state $E(i, j)$}
\nomenclature{$Q$}{Transition rate matrix.}
\nomenclature{$\phi$}{Configuration matrix.}
\nomenclature{$\Upsilon$}{Count of failures occurring.}
\nomenclature{$p_a$}{Probability of an arrival request.}
\nomenclature{$p_m$}{Probability of a mined block.}
\nomenclature{$p_r$}{Probability of a rejected request.}
\nomenclature{$p_s$}{Probability of a serviced request.}
\nomenclature{$T_a$}{Average time between arrivals requests.}
\nomenclature{$T_m$}{Average time between mined blocks.}
\nomenclature{$T_r$}{Average time between rejected requests.}
\nomenclature{$T_s$}{Average time between serviced requests.}
\printnomenclature

\section{B-RAN USAGE SCENARIOS} \label{S:bran scenarios}
B-RAN enables collaboration among different service providers and users in a secure, private, and dependable manner. This is achieved by blending blockchain with virtualization, multiple-access edge computing (MEC), and artificial intelligence (AI) functionalities. B-RAN opens the door to a number of attractive usage scenarios, such as fronthaul network of fixed topology, advanced coverage expansion, and advanced connectivity of mobile nodes. In the rest of the section, we document the aforementioned scenarios.

\subsection{Fronthaul network of fixed topology}
\color{black}
In the fronthaul network of fixed topology, each user equipment (UE) performs tasks that demand significant computing power and time sensitivity, such as navigation, video streaming, or virtual reality. It is assumed that the base stations (BSs), which may belong to different service providers, are equipped with MEC capabilities. BSs have computing resources and can carry out AI tasks. UEs with resources can offload tasks to various edge infrastructures using resource allocation strategies, such as those optimized for cooperative transmission in C-RAN environments~\cite{Yang2021}. Additionally, coordinated multi-point (CoMP) connectivity, managed through joint precoding and resource allocation strategies, can improve system reliability and energy efficiency in scenarios involving multiple BSs~\cite{Jiang2023}. The need for B-RAN stems from the absence of trust-based interactions between UEs and BSs.
\color{black}

\begin{figure*}
    \centering
    \includegraphics[width=1\linewidth]{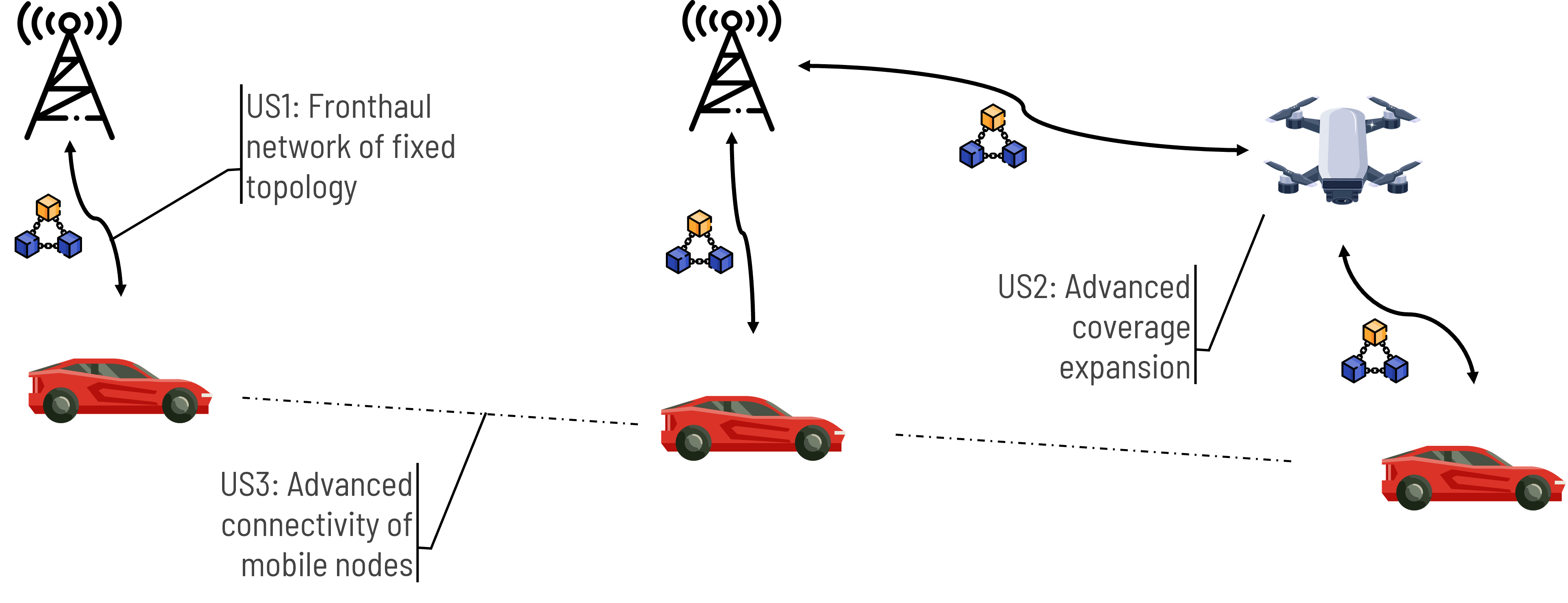}
    \caption{Network topology.}
    \label{Fig:topology}
\end{figure*}

\subsection{Advanced coverage expansion}
\color{black}
The concept of advanced coverage expansion plays significant role in future communication networks in order to address the increasing demand for reliable high speed connectivity in various environments. It involves utilizing state-of-the-art approaches, like using infrastructure as relay nodes~\cite{Bulakci2011}, implementing node structures, and employing efficient connectivity models to enhance network performance~\cite{Vila2024}. These methods not only ensure expanded and higher quality network coverage but also significantly improve energy efficiency. B-RAN is envisioned to augment security in coverage expansion scenarios through the integration of blockchain to strengthen confidentiality and trust in communications. By incorporating these elements, B-RAN is capable to meet the complex and evolving needs of modern connectivity, while also emphasizing the importance of adaptability, security, and efficiency in network expansion.
\color{black}

\subsection{Advanced connectivity of mobile nodes}
\color{black}
This scenario promotes communication between vehicles and BSs as well as between different vehicles. Information can be shared among vehicles, where one vehicle acts as the intermediate node to forward the data to the BS. Establishing a network that supports both vehicle-to-vehicle (V2V)~\cite{Tokarz2020} and vehicle-to-BS communication is necessary~\cite{Malik2020},~\cite{Hua2019}. In addition, an important goal is to ensure wide coverage and efficiency in terms of latency, throughput, and energy consumption. 

Although, vehicular communications rely on ground-based infrastructure for V2V transmission, the growing demands of services call for more stringent requirements that these traditional methods cannot fulfill. For instance, when vehicles travel far from the BS, ensuring latency for real-time applications requires synchronization among vehicles and becomes challenging if communication must go through the BS. V2V communications address this by enabling direct information exchange without relying on centralized infrastructure. 

While eliminating the reliance to the BS can offer advantages, it also comes with drawbacks such as lack of a centralized entity responsible for network security management. B-RAN aims to ensure security and privacy by using pseudonyms when sharing data since trust cannot be assumed. Moreover, B-RAN can enable support for multi-hop communications among vehicles to minimize connectivity gaps and extend the coverage of the network.
\color{black}

\section{Network topology} \label{S:system_model}
The realization of the aforementioned usage scenarios is founded upon not only point-to-point (p2p) but also multi-hop connectivity. If we take for instance a mobile connectivity use case of a vehicle moving inside the coverage area of the network, p2p connectivity can suffice for providing network services to the UE. However, when the vehicle reaches the limits of the network's coverage area, p2p links can no longer provide adequate quality for the service. In this case, an ad-hoc network must be instantiated by a node that is located close to the UE and is capable of extending the network coverage and providing connectivity. This intermediate node is connected at the same time to the mobile UE and the BS; thus, providing the service though multi-hop connectivity.

This high-mobility use case is depicted in~Fig.~\ref{Fig:topology}, which illustrates the different B-RAN usage scenarios as a vehicle moves through the network and eventually exits its coverage to be served by the intermediate node. Specifically, the fronthaul network of fixed topology usage scenario is applicable to direct connectivity cases inside the coverage area of the network, the advanced connectivity of mobile nodes describes the entire movement of the vehicle, and the advanced coverage expansion scenario is applicable after the vehicle moves beyond the limits of the fixed infrastructure and an ad-hoc network is deployed by the intermediate node.

In this paper, we model the B-RAN network dynamics through Markov-chain theory. To achieve this, we split the theoretical modeling in two network topologies. The first assumes a direct connectivity scenario with the UE connected directly to the BS; thus, providing the service through the primary blockchain of the network. \color{black} Of note, there are three main blockchain architectures i.e. public, private and consortium, that can be used in an implementation~\cite{Zheng2018}. Each blockchain type has different characteristics, specification, and requirements; however, the flexible and adaptable nature of the presented model makes it applicable in all possible scenarios regardless of the blockchain type. \color{black} The second topology tackles the multi-hop connectivity case of advanced coverage expansion scenario; thus requires the establishment of a hierarchical B-RAN architecture that deploys a secondary blockchain between the intermediate node and the UE in order to provide the service outside of the network's primary coverage area. Alongside the blockchain architecture, it is essential to mention the consensus mechanisms. According to relevant studies~\cite{Ahn2024},~\cite{Zhou2023} the consensus algorithms can be categorized into a variant of types, with the most used to be Proof of Work(PoW), Proof of Stake (PoS) and Byzantine Fault Tolerance (BFT) each of them with unique attributes. For example, a characteristic feature of PoW is the fact that is resource intensive, as it requires nodes to solve difficult problems, a feature that offers strong security level but its highly demanding on energy consumption. On the other hand, PoS completes its role based on validator's stakes, thus consuming less energy. Finally, BFT algorithm is ideal for distributed models where the fault tolerance has an important role in the overall model. In conclusion, we can distinguish that each consensus mechanism has its strengths and weaknesses by trying to satisfy both the overall speed of the model and its security. The ideal choice at the end, it depends on the blockchain model and the attributes (e.g. security, latency times, energy consumption) that would like to address the most at every scenario. The proposed framework can model different deployments of blockchain through the employed Markov-chain theory-based approach that depends on the probabilistic behavior of the blockchain. \color{black}
\begin{figure*}
    \centering
    \includegraphics[width=1\linewidth]{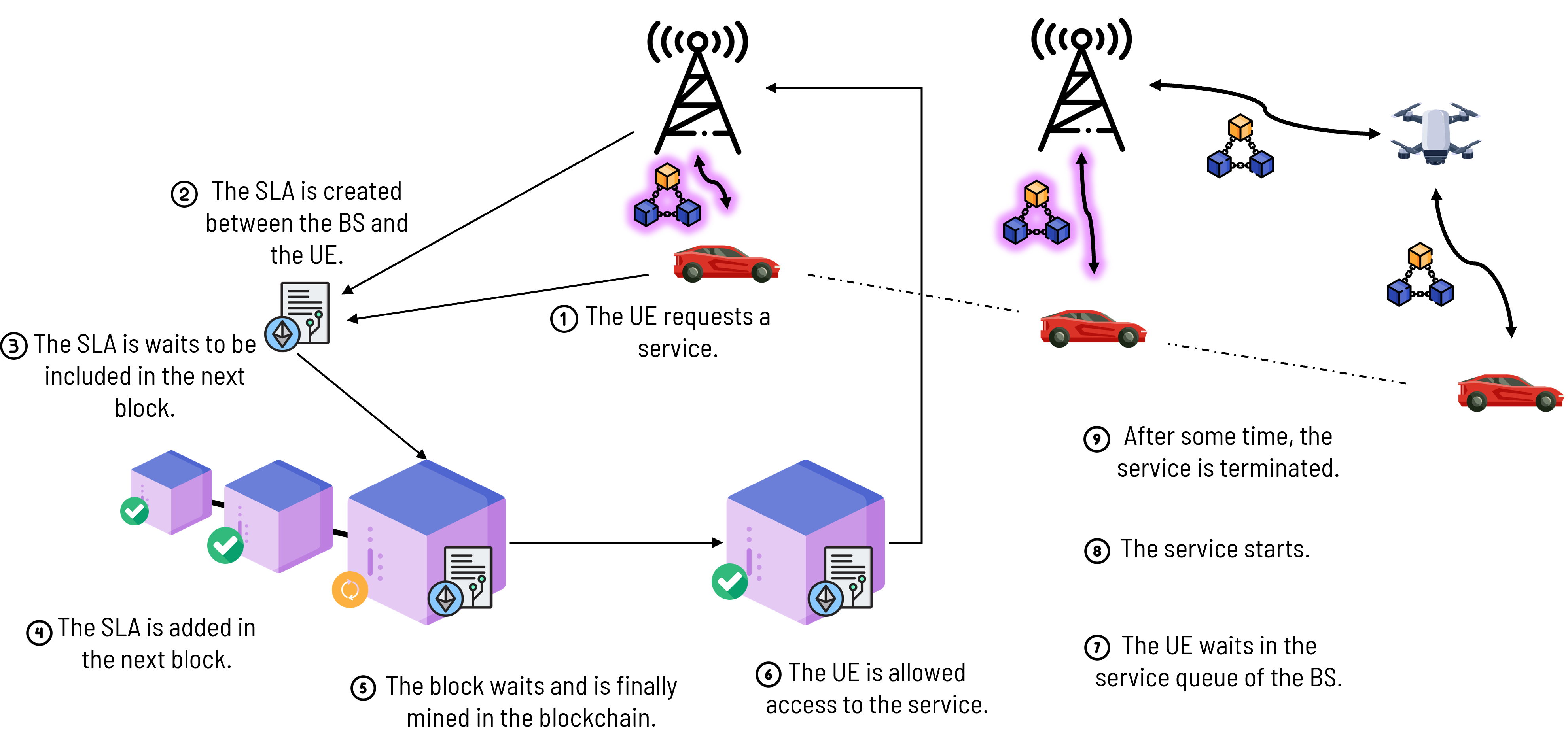}
    \caption{B-RAN architecture.}
    \label{Fig:blockchainSingle}
\end{figure*}

\section{B-RAN Modeling} \label{S:bran_modeling}
In this section, we present the B-RAN model. We employ a Markov chain model to capture the probabilistic transitions between different system states and delve into their dynamic behavior during the operations of B-RAN. Additionally, we introduce an extension of the single-chain B-RAN model, which is characterized by nested blockchains that create a HB-RAN architecture. This HB-RAN model provides an novel solution for assessing the performance of coverage expansion scenarios, like ad-hoc deployments and cell-free network access in terms of security and reliability. Through this approach, our aim is to illustrate the operational dynamics of B-RAN, while offering essential insights for its optimization and improvement.

The B-RAN model is depicted in~Fig.~\ref{Fig:blockchainSingle} and illustrates its operation through the utilization of two queues. The first queue models service requests that wait to be include in a blockchain block, while the second queue models confirmed requests that wait to be serviced by the network. In more detail, the first queue operates based on the principles of a $M/M/1$ queue, in which requests arrive with a Poisson distribution with rate $R_a$, and their processing times are governed by memoryless exponential distributions with a rate of $R_m$. \color{black} The Poisson process is ideal to model network traffic since the aggregation of multiple i.i.d. processes tends to a Poisson process for a sufficient number of events. Also, its simplicity and mathematical tractability enable efficient simulation and analysis, making it straightforward to generate and evaluate traffic patterns. In addition, the superposition and splitting properties allow complex network scenarios-such as multiplexed traffic or routed paths-to be modeled with minimal computational overhead. \color{black} Each of the request is processed within blockchain blocks that may contain a maximum of $k$ number of requests per block. Additionally, the second queue is modeled as a $M/M/s$ queue, with $s$ representing the maximum number of access links. Requests arrive based on a Poisson distribution, while their processing times are characterized by memoryless exponential distributions. Based on the above model, we assume that, at any given time $t$, the system's state can be summarized by using the mathematical expectation for every state, $E[i, j]$, where $E$  represents the state, $i$ denotes requests pending mining into a block, and $j$ stands for requests waiting to be served. The complex interaction between queue dynamics are highlighted through the aforementioned approach, capturing the crucial states of pending requests on their way to block inclusion and service.

\begin{figure}
    \centering
    \includegraphics[width=1\linewidth]{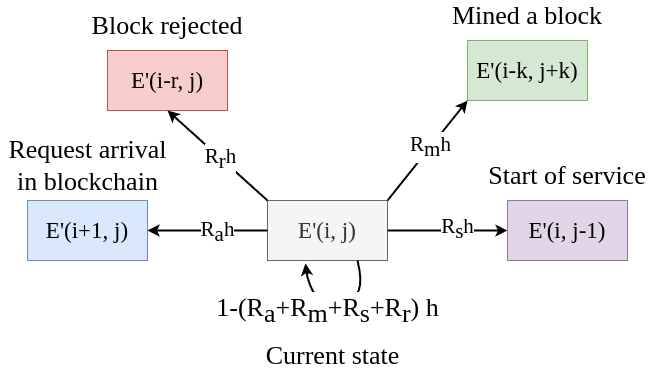}
    \caption{B-RAN Markov chain model.}
    \label{Fig:MC}
\end{figure}

\begin{figure*}
    \centering
    \includegraphics[width=1\linewidth]{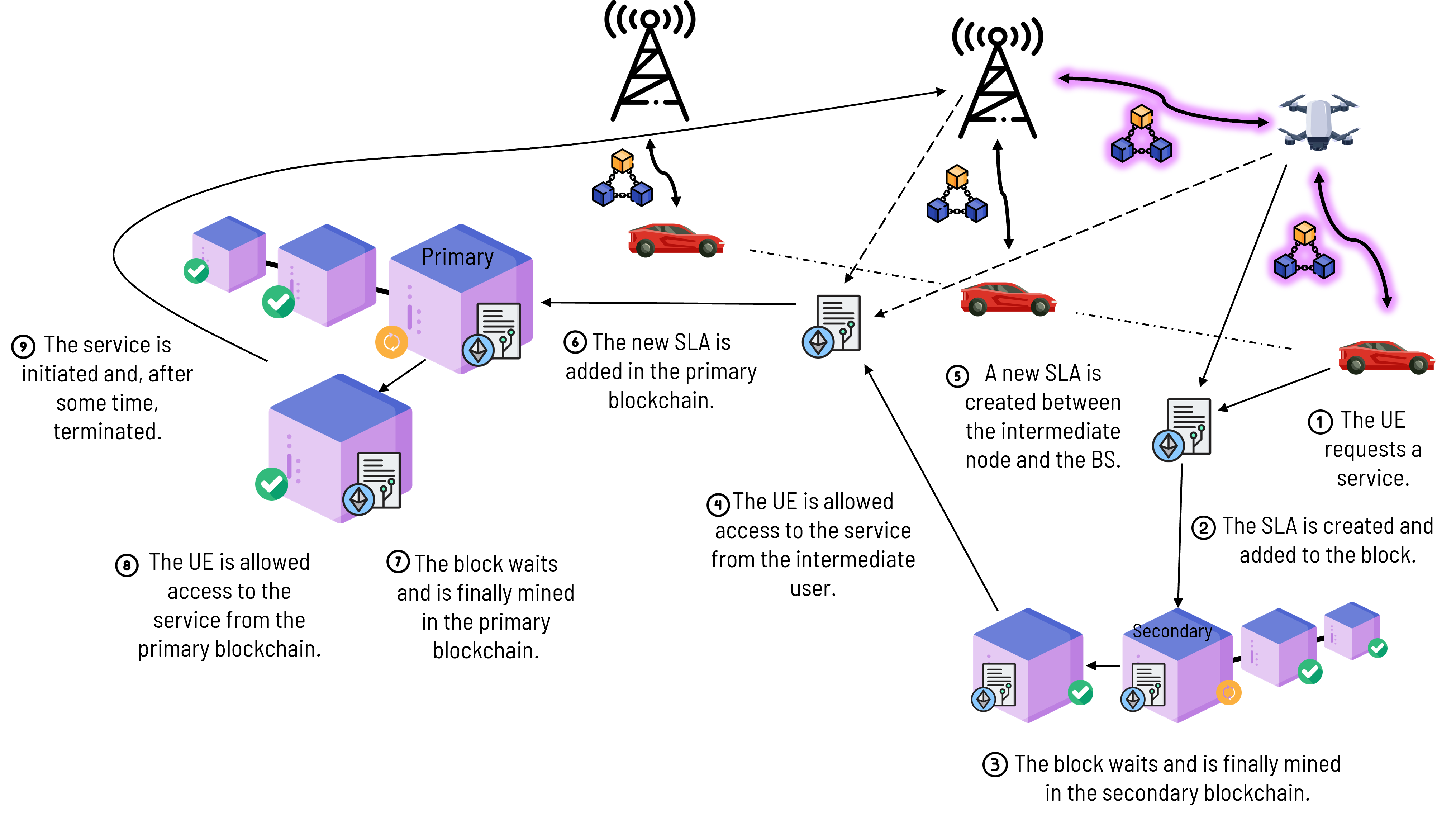}
    \caption{Hierarchical blockchain architecture.}
    \label{Fig:arch}
\end{figure*}

\subsection{Markov chain model}
The possible states can be aptly portrayed as a continuous time-homogeneous Markov process; thus, embodying all the defining characteristics inherent to a Markov chain~\cite{Ling2021}. As depicted in~Fig.~\ref{Fig:MC}, the Markov chain model is defined by its current state, $E[i, j]$, at time $t$, and five discrete states that capture various configurations of the system. The transitions between these states take place over minimal time interval $h\to 0$.

When a new request is received, the next state is denoted by $E'[i + 1, j]$ and signifies an increase in the number of pending requests for blockchain, $i$, as an additional request is added to the next block. Of note, in this case, the number of requests waiting service, $j$, remains unchangeable. This reflects the fact that only one event can take place at any given $h$. The probability of this transition is defined~as 
\begin{align}
p_a = R_a \, h,
\label{Eq:pa}
\end{align}
where $R_a$ stands for the rate at which an arrival request~occurs. 

Next, the transition to $E'[i-k, j+k]$ describes the case in which  a block is successfully mined. The probability of successful block mining can be expressed as 
\begin{align}
p_m = R_m \, h,
\end{align}
with $R_m$ being the mining rate of a block.
Note that $p_m$ depends on $k$, which denotes the maximum number of requests that can be included in a single block. A successful block mining event is related on the number of pending requests and the block size number. If the number of pending requests, $i$, is equal to or less than the threshold $k$, then all pending requests are successfully mined in a single block; thus, increasing $j$ by $i$; the subsequent state is denoted by $E'[0, j + i]$. Conversely, if the number of pending requests surpasses the threshold $k$, only a maximum of $k$ requests can be mined, while the remaining requests remain pending; in this case, the next state can be written as $E'[0, j + k]$.

The transition to the $E'[i, j - 1]$ state models the service of a request and is associated with a probability 
\begin{align}
p_s = R_s \, h,
\end{align}
where $R_s$ is the service rate.  
This transition indicates a reduction of $1$ in the $j$ queue, signifying the commencement of service for the corresponding request. Note that the number of pending requests, $i$, remains unaffected, as service initiation influences only the queue of blocks and not the pending requests.

Additionally, the transition to state $E'[i-r, j]$ characterizes the rejection of a request due to factors like authentication problems, insufficient resources, and so on. In this case, $r$ represents the number of rejected requests. This transition is governed by the rejection probability, which is defined~as
\begin{align}
p_r=R_r \, h,
\label{Eq:pr}
\end{align}
with $R_r$ standing for the rejection rate. 
When a rejection event occurs, $i$ decreases by $r$ since the block that contains the rejected request is discarded and the remaining requests need to be included in the next block. Meanwhile, $j$ remains unchanged emphasizing that the rejected block did not advance to the mining stage.

Finally, there is a  probability that none of the aforementioned events occur; this signifies the idle state. The idle state is denoted by $E'[i, j]$ with its probability being written as 
\begin{align}
    p_i = 1-\left(p_a + p_m + p_s + p_r\right), 
\end{align}
or, after applying~\eqref{Eq:pa}--\eqref{Eq:pr}, 
\begin{align}
p_i = 1 -\left( R_a + R_m + R_s + R_r \right) \, h.
\end{align}
The idle state represents that the system remains unchanged at the given time without moving to any available states. The probability of this scenario captures the possibility of no requests coming in, no requests being rejected, no mining successes, and no service operations being completed.

\subsection{Hierarchical B-RAN model}
\color{black}
In order to support the  scenarios, which were documented~in~Section~\ref{S:bran scenarios}, HB-RAN deployments are required. In the scenario of advanced coverage expansion, the end user is unable to directly connect to the BS of its internet service provider. However, it can establish a direct link with an intermediate user (IU) that is already connected to the BS via the primary blockchain. Therefore, a secondary blockchain is created between the intermediate and the end user to ensure security, privacy, and trust, as seen in Fig.~\ref{Fig:arch}. 

This procedure generates a smart contract between the end user and the IU. Consequently, an end-user request from the secondary network, as illustrated in Step 1 of the figure, is initiated to establish a connection and create a Service Level Agreement (SLA) in Step 2, after which it is inserted into the secondary blockchain. At first, this request enters a $M/M/1$ queue and waits to be included in a block; this process is depicted in Step 2 of the figure. Next, the mining phase begins  by the blockchain network in order to verify the request. After a successful mining process, the request is forwarded to a second queue, where blocks are waiting to be serviced using a multiple-server queuing model~$M/M/s$. 

Once the request is validated through the secondary blockchain (Step 4), a corresponding request is formed in the primary blockchain. Upon accessing the primary blockchain by a new SLA thas has been created from the secondary blockchain, as depicted in step 5, it joins the $M/M/1$ queue of the primary blockchain (Step 6) for block inclusion and then waits for $N$ confirmations to validate the block, according to step 7 of the figure. Afterwards, it transitions into a distinct stage that waits in the $M/M/s$ queue of the primary blockchain to begin its service (Step 8). Once the request's service starts on the main blockchain, it switches to the secondary blockchain, and its end-to-end (e2e) latency can be measured, reflecting the initiation of service as depicted in Step 9. This refers to the total time spent navigating across both primary and secondary blockchains.
\color{black}

\section{PERFORMANCE EVALUATION} \label{S:performance_evaluation}
In this section we analyze the performance of the BRAN framework, concentrating on assessing the vulnerabilities and estimating the system latency. We provide important insights into the fundamental concepts of B-RAN by mathematically modelling latency. Moreover, we consider the scenario of a double spending attack on BRAN and provide closed form expression of the probability of successful attack. This part demonstrates, by careful inspection and analysis, the robustness and advantages of the B-RAN model across a range of network topologies and with potential security risks.

\subsection{Latency}
The proposed framework utilizes two queues to model the complex dynamics of incoming requests and their processing in blockchain blocks. As explained earlier, a $M/M/s$ queue simulates the latency caused by service initiation and processing, while a $M/M/1$ queue handles requests that are waiting to be included in the blockchain. The end-to-end latency of the system is a result of both queues. The expected value of the waiting time due to the $M/M/1$ queue in the B-RAN model can be analytically expressed as in~\cite{Cooper1981}
\begin{align}
    \tau_1 = \frac{1}{R_m - R_a} ,
\end{align}
where $R_a$ stands for arrival rate and $R_m$ for service rate. Moreover, the latency generated by the $M/M/s$ queue can be written as in~\cite{Kelton1985}
\begin{align}
    \tau_2 = \frac{C(s, \frac{R_a}{R_s})}{sR_s - R_a} + \frac{1}{R_s} ,
\end{align}
with the first term's nominator expressing the Erlang C formula, which depends on $s$, $R_a$, and $R_m$. Furthermore, the confirmation process of the blockchain also creates some additional delay that can be calculated as
\begin{align} 
    \tau_3 = \frac{N-1}{R_m}, 
\end{align}
where $N$ denotes the number of confirmations and $R_m$ represents the block generation rate. At this point, Little's Law has been applied to establish a relationship between the expected latency and the queue length. Little's Law asserts that the arrival rate multiplied by the average time an item spends in the system equals the average number of transactions in a stable system. Consequently, the expected sojourn time, $\tau_s$, which quantifies how long each service request remains within its specific system state, can be expressed as
\begin{align} 
    \tau_s = \tau_1 + \tau_2 + \tau_3 .
\end{align}

As a result, the average latency of B-RAN, $\tau_t$, can be evaluated as
\begin{align} 
    \tau_t = \tau_s -\frac{1}{R_s}.
\end{align}
It is important to highlight that, in the HB-RAN model, the same process is followed for the evaluating the latency. This method calculates the time it takes for a request to be served by the primary blockchain. However, in the coverage expansion scenario, the end-to-end latency is measured by combining the delays incurred by both the primary and secondary blockchains. 

Based on the aforementioned, we now investigate the B-RAN service latency for the single confirmation scenario. In this scenario, the current state is expressed as $E\left(i, j\right)$ with $i$ and $j$ denoting the pending requests awaiting assembly into a block and the confirmed requests ready for service, respectively. Let $P_{i,j} (t) = P\{X(t) = E(i, j)\}$ denote the probability of the queue being in state $E(i, j)$ at time $t$. Additionally, we assume that the transition probability $\operatorname{P}\{X(t)=$ $\left.E \mid X(t+h)=E^{\prime}\right\}$ characterizes the queuing model. All transition probabilities are zero except for events of arrivals, mined blocks, rejected blocks, or start of service. \color{black} The non zero probabilities of the system are given by 
\begin{align}
    \begin{split}
        &\operatorname{P}\left\{E\left(i, j\right) \mid E^{\prime}\left(i, j\right)\right\} \\ &=\frac{R_a}{\!R_a\!+\!R_m\!+\!R_r\!+\!R_s^j}\left(\left(\!R_a\!+\!R_m\!+\!R_r\!+\!R_s^j\right) h+o(h)\right.
    \end{split}
\end{align}
In case a new request arrives in the system, its probability is given by
\begin{align}
\lim _{h \rightarrow 0} 
\operatorname{P}\left\{E\left(i, j\right) \mid E^{\prime}\left(i+1, j\right)\right\}=R_a h+o(h).
\end{align}
Additionally, when a block gets mined, the probability of this event can be written as
\begin{align}
& \lim _{h \rightarrow 0} 
\operatorname{P}\left\{E\left(i, j\right) \mid E^{\prime}\left(i-k, j+k\right)\right\}=R_m h+o(h).
\end{align}
In case a contract gets rejected from the block, its probability can be expressed as
\begin{align}
\lim _{h \rightarrow 0} 
\operatorname{P}\left\{E\left(i,j\right) \mid E^{\prime}\left(i - r, j\right)\right\}=R_r h+o(h). 
\end{align}
In case a contract gets serviced, the probability of this event is given by
\begin{align}
\lim _{h \rightarrow 0}
\operatorname{P}\left\{E\left(i, j\right) \mid E^{\prime}\left(i, j-1\right)\right\}=R_s^j h+o(h),
\end{align}
where \( o(h) \) represents the probability that more than one event occurs at time \( t \) as \( h \to 0 \).

\begin{figure}
    \centering
    \includegraphics[width=0.8\linewidth]{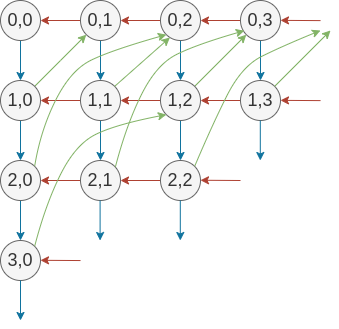}
    \caption{State space scenario for $k = 2$ and $r = 1$}
    \label{Fig:fig5}
\end{figure}

\color{black}
The sum of all transition probabilities should be equal to unity, which can be expressed~as
\begin{align}
    \begin{split}
        P_{i, j}&(t+h)-P_{i, j}(t) = P_{i-1, j}(t) R_a h+P_{i, j+1}(t) R_s^{j+1} h \\
        &\quad -P_{i, j}(t)\left(R_a+R_m+R_s^j+R_r\right) h,
    \end{split}
    \label{eq:1}
\end{align}
where $R^s$ is the completion rate and it's defined by $R_s^j = min(j,s)R_s$ for $0\leq j\leq s$, since at most $s$ can be in service at the same time. By assuming that $h \rightarrow 0$, we obtain the steady-state distribution of B-RAN~as
\begin{align}
    P_{i-1, j} R_a+P_{i, j+1} R_s^{j+1}-P_{i, j}\left(R_a+R_m+R_s+R_r\right)=0 .
    \label{eq:3}
\end{align}
where $\frac{d}{dt} P_{i, j}(t)=0$. Specifically, in the boundary case of $(i=0)$,~\eqref{eq:3} can be rewritten~as
\begin{align}
    \begin{split}
    & \left(\sum_{\ell=1}^j P_{\ell, j-\ell}\right) R_m+P_0^{j+1} R_s^{j+1} \\
    & \quad-P_{0, j}\left(R_a+R_s^j+R_r\right)=0, \quad \forall j=0,1,2, \ldots \\
    \end{split}
    \label{eq:4}
\end{align}
where
\begin{align}
    \begin{split}
        P_{0,1} R_s^1 - P_{0,0} R_a &= 0.
    \label{eq:4_2}
    \end{split}
\end{align}
The differential-difference equations~\eqref{eq:3}-~\eqref{eq:4_2} are the forward Kolmogorov equations~\cite{kleinrock1976}, which can be rewritten more concisely in a probability vector given~by 
\begin{equation}
    \mathbf{P}=\left[\begin{array}{lllllll}
    P_{0,0} \mid & P_{1,0} & P_{0,1} \mid & P_{2,0} & P_{1,1} & P_{0,2} & \cdots
    \end{array}\right]^T ,
    \label{eq:5}
\end{equation}
or in matrix form~as
\begin{equation}
    \mathbf{Q P}=\mathbf{0},
    \label{eq:qw0}
\end{equation}
with $\mathbf{Q}$ being the infinitesimal generator or transition rate matrix. Each entry in $\mathbf{Q}$ equals the corresponding transition rate given by $\frac{d}{d h} \operatorname{Pr}\left\{X(t)=E \mid X(t+h)=E^{\prime}\right\}$, depending solely on the B-RAN configuration tuple $\Phi=\left\{R_a, R_m, R_r, R_s, s\right\}$. It can be numerically calculated by utilizing the sum probability condition, $\mathbf{1}^T \mathbf{P}=1$~as
\begin{align}
    \left[\begin{array}{c}
    \mathbf{Q} \\
    \mathbf{1}^T
    \end{array}\right] \mathbf{P}=\left[\begin{array}{l}
    \mathbf{0} \\
    1
    \end{array}\right] .
    \label{eq:6}
\end{align}
The state transition relationships that can be calculated based on the presented analysis for the one confirmation case are presented in~Fig.~\ref{Fig:fig5}.

From~\eqref{eq:6}, the steady-state distribution, $\mathbf{w}(\Phi)$, can be analyzed as an implicit function of $\Phi$. Of note the waiting space of B-RAN has no maximum limit, resulting in infinite dimensions for the vector $\mathbf{w}$. In numerical calculations, we approximate the infinite-dimension solution with sufficiently large but finite dimensions as, in practice, the number of UEs cannot be infinite. Thus, the aggressive load $R_a$ must be less than $R_s$ for stability. In order To analyze the average latency in B-RAN, we consider the limiting distribution $\mathbf{w}(\Phi)$ to obtain the average number of waiting requests $N(\Phi)$~as
\begin{align}
    \mathbb{E}\{N(\Phi)\} = \sum_{i, j}(i+j) \cdot P_{i, j}(\Phi) .
    \label{eq:7}
\end{align}
Applying Little's Law, we can link the expected queue length and average latency as the expected sojourn time, $\mathrm{L}_s(N, \Phi)$, includes both waiting and service latency and can be expressed~as
\begin{align}
    \begin{split}
        \mathrm{L}_s(N=1, \Phi) & = \mathbb{E}\{\mathrm{N}(\Phi)\} / R_a \\
        & = T_a \sum_{i, j}(i+j) P_{i, j}(\Phi) ,
    \end{split}
    \label{eq:8}
\end{align}
and, therefore, the average latency for the one-confirmation scenario can be expressed based on the limiting distribution in~Eq.~\eqref{eq:qw0}~as
\begin{align}
    \mathrm{L}(N=1, \Phi) = T_a \sum_{i, j}(i+j) P_{i, j}(\Phi) - T_s ,
    \label{eq:9}
\end{align}
with $T_s$ denoting the service time. 

So far, only the one-confirmation scenario was considered. When investigating the generic $N$-confirmation problem, the huge number of variables in eq.~\ref{eq:qw0} makes analysis of a queue with $(N+1)$-dimensional state space difficult. However, if we consider that once a request is included into a block it has to wait for $N-1$ confirmations after the initial confirmation, eq.~\eqref{eq:9} can be rewritten~as
\begin{align}
    \begin{split}
    \mathrm{L}(N, \Phi) & = \mathrm{L}(1, \Phi) + \mathbb{E}\left\{\sum_{n=2}^N U_n^m\right\} \\
    & = T_a \sum_{i, j}(i+j) P_{i, j}(\Phi) + T_m(N-1) - T_s
    \end{split}
    \label{eq:10}
\end{align}

\begin{figure*}
    \begin{align}
        Q = \left[
        \resizebox{0.87\linewidth}{!}{
        \begin{tabular}{c|cc|ccc|cccc}             
             $-R_a$  &  $R_r$  &  $R_s$  &  &  &  & & & & \\
             \hline
             $R_a$ & $-(R_a\!\!+\!\!R_m\!\!+\!\!R_r)$  &  & $R_r$ & $R_s$ &  &  &  & &\\     
             & $R_m$  & $-(R_a\!\!+\!\!R_s)$ &  & $R_r$ & $R_s$ &  &  & &\\
             \hline
             &  $R_a$  &  & $-(R_a\!\!+\!\!R_m\!\!+\!\!R_r)$ &  &  & $R_r$ & $R_s$ &  &\\    
             &  &  $R_a$& & $-(R_a\!\!+\!\!R_m\!\!+\!\!R_r\!\!+\!\!R_s)$ &  &  & $R_r$ & $R_s$ & \\             
             &  &  & $R_m$ & $R_m$ & $-(R_a\!\!+\!\!R_s)$ &  &  & $R_r$ & $R_s$\\
             \hline
             &  &  & $R_a$ &  &  & $-(R_a\!\!+\!\!R_m\!\!+\!\!R_r)$ &  &  & \\             
             &  &  &  & $R_a$ &  &  & $-(R_a\!\!+\!\!R_m\!\!+\!\!R_r\!\!+\!\!R_s)$ &  & \\             
             &  &  &  &  & $R_a$ & $R_m$ &  & $-(R_a\!\!+\!\!R_m\!\!+\!\!R_r\!\!+\!\!R_s)$ & \\      
             &  &  &  &  &  &  & $R_m$ & $R_m$ & $-(R_a\!\!+\!\!R_s)$ 
        \end{tabular} }
        \right] 
        \label{Eq:Qmatrix}
    \end{align}
\end{figure*}
Based on~\eqref{eq:10}, the transition rate matrix, $Q$, becomes an infinitely dimensional structure that encapsulating all possible rates for every state and is given~by~\eqref{Eq:Qmatrix}. It visualizes the rate at which changes between states occur in a stochastic process. Each element of the matrix represents the rate of transitioning from one state to another. The off-diagonal elements represent the rates of shifting between separate states, while the diagonal elements express the rate of the idle state and ensure that the sum of all elements in a row equals to zero. This matrix is crucial for the calculation of the latency since it determines the system's behavior throughout the stochastic system. 

If we take the initial state $E[0,0]$ for example, which denotes no pending requests to be mined and no requests awaiting servicing, according to~\eqref{Eq:Qmatrix}, we have two possible transitions. Either the system stays idle, or a new request arrives with a rate of $R_a$. Representing the absence of further transitions, the rate for the system to remain inactive is calculated as the negative of the arrival rate $R_a$. As we get into more complicated stages, the number of possible outcomes grows. For each state, their future possible states can be identified through their corresponding transition rates. For instance, from state $E[0,1]$, there are three possible future states, specifically
\begin{enumerate}
    \item A service is completed at a rate of $R_s$ and the state changes into the state $E^{\prime}[0,0]$  as the number of requests decreases by one.
    \item A new request arrives at a rate of $R_a$, transitioning to state $E^{\prime}[1,1]$. 
    \item The system remains idle, with this rate equal to the negative sum of the arrival and service rates $-(R_a + R_s)$. 
\end{enumerate}
All in all, by analyzing the transition rates in the Q matrix, we can accurately predict all the possible future states from any given state. 

\begin{figure*}
    \centering
    \includegraphics[width=1\linewidth]{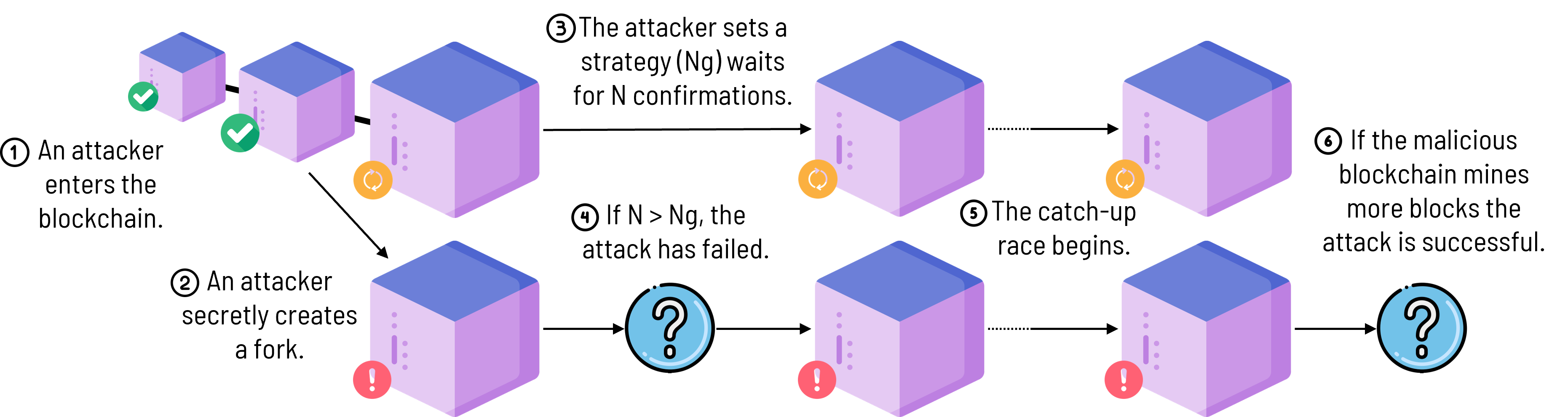}
    \caption{Procedural illustration of an alternate history attack in B-RAN.}
    \label{Fig:attack_diagram}
\end{figure*}

\subsection{Security} \label{S:Attack}
\color{black}
The incorporation of blockchain technology into RAN systems has the potential to improve security and avoid attacks by malicious users. The decentralized and transparent design of blockchain enhances its resilience against attacks. However, the structure of blockchain raises new security risks that were not present in earlier RAN systems. A typical example is the alternative history attack, which includes malicious attempts to modify the transactions in the blockchain's history. This attack scenario is examined in detail in the following section, along with how it could affect B-RAN's performance or compromise its dependability and security.

In the case of the alternative history attack, as seen in Fig.~\ref{Fig:attack_diagram}, an attacker initially gains access to the blockchain as a regular user. At some point, along with the official mining process, the attacker creates an exact duplicate of the official blockchain. Despite the differences in mining rates between the two versions (official and malicious), official blockchain activities are unaffected. The mining rate of the malicious fork, $R_m$, is determined by the computational capabilities of the attacker. Additionally, the symbol $beta$ represents the ratio between legitimate and malicious blockchains. Once the tampered block gathers $N$ confirmations, the attacker initiates a mining race to catch up with the official blockchain. The attacker evaluates the length of the malicious fork compared to the original chain. If this difference falls below a specified threshold $N_g$, the attacker persists in mining until the malicious chain surpasses the official one and deems the attack as successful. On the contrary, if the difference exceeds $N_g$, the attacker ceases the attack deeming it unsuccessful. The likelihood of a successful alternative history attack depends on the attacker's relative mining rate, $\beta$, the required number of confirmations, $N$, and the attacker's strategy, $N_g$. 

To evaluate the probability of a successful attack we assume stable strategy level and mining rates. We consider a scenario where the probability of extending the official chain by one block is $\frac{1}{1+\beta}$, while the likelihood of an attacker to find the next block is $\frac{\beta}{1+\beta}$. This implies that the mining process can be modeled by a series of independent Bernoulli trials with a success probability of $\frac{1}{1+\beta}$. For the attack to be successful, the attacker must deliberately wait for $N$ confirmations. At the same time, the attacker generates $n_Y$ blocks on the malicious fork. Consequently, the stochastic variable denoting the number of failures, $Y$ and follows a negative binomial distribution, $Y \sim \mathcal{NB}(N, 1 /(1+\beta))$, with the probability mass function given~by
\color{black}
\begin{align}
    \begin{split}
        \operatorname{Pr}&\left\{Y=n_Y ; N, \frac{1}{1+\beta}\right\} = \\
        & = \binom{n_Y+N-1}{n_Y}\left(\frac{1}{1+\beta}\right)^N\left(\frac{\beta}{1+\beta}\right)^{n_Y}.
    \end{split}
\end{align}

Afterwards, both the malicious and the official blockchains start mining with the attacker trying to outperform the official network. If this happens, the attacker can publish the malicious chain and rewrite the confirmed history. However, if the fraudulent chain lags behind by $N_g$ blocks, the attacker abandons the attempt. Let $P_n=\operatorname{Pr}{\{\text{Win} \mid z=n\}}$ denote the probability of the attacker winning despite starting with a delay of $n$ blocks. Two special cases become evident, specifically $P_{-1}=1$ and $P_{N_g}=0$. If the attacker finds the next block, the malicious chain shortens by $n-1$ blocks compared to the benign chain, and the success probability becomes $P_{n-1}$. Conversely, if the official blockchain mines a block, the attacker falls further behind to $n+1$ blocks and the success probability decreases to $P_{n+1}$. By conditioning on the outcome of the first generated block, the probability of the attacker winning can be written~as 
\begin{align}
    P_n=\frac{1}{1+\beta} P_{n+1}+\frac{\beta}{1+\beta} P_{n-1}, \quad 0 \leq n<N_g,
    \label{Eq:Pn}
\end{align}
which can be further reformulated~as
\begin{align}
    P_{n-1}-P_n=\frac{1}{\beta}\left(P_n-P_{n+1}\right), \quad 0 \leq n<N_g.
    \label{Eq:P_n-1}
\end{align}
\color{black}
For $n=N_g-1$,~\eqref{Eq:P_n-1} can be rewritten~as
\begin{align}
    P_{N_g-2}-P_{N_g-1}=\frac{1}{\beta}\left(P_{N_g-1}-P_{N_g}\right)=\frac{1}{\beta} P_{N_g-1},
\end{align}
which, through recursion, yields
\begin{align}
    P_{N_g-n-1}-P_{N_g-n}=\frac{1}{\beta^n} P_{N_g-1}, \quad 0 \leq n<N_g,
\end{align}
that can be rewritten as
\begin{align}
    P_{N_g-n-1} & =P_{N_g-1}+\sum_{m=1}^n \frac{1}{\beta^m} P_{N_g-1} .
    \label{Eq:P_N_n-1_1}
\end{align}
By expanding the sum,~\eqref{Eq:P_N_n-1_1} can be transformed into 
\begin{align}
    P_{N_g-n-1} = \begin{cases}P_{N_g-1} \frac{1-1 / \beta^{n+1}}{1-1 / \beta}, & \text { if } \beta \neq 1 \\
    P_{N_g-1}(n+1), & \text { if } \beta=1 .\end{cases}
    \label{Eq:P_Ng_n-1_2}
\end{align}
Next, by utilizing the boundary condition $P_{-1}=1$,~\eqref{Eq:P_Ng_n-1_2} can be rewritten~as
\begin{align}
    P_{N_g-1} = \begin{cases}\frac{1-1 / \beta}{1-1 / \beta^{N+1}} & \text { if } \beta \neq 1 \\
    \frac{1}{N_g+1} & \text { if } \beta=1 \end{cases}.
\end{align}
Hence, we derive the expression for $P_n$~as
\begin{align}
    P_n= \begin{cases}\frac{\beta^{n+1}-\beta^{N_g+1}}{1-\beta^{N+1}} & \text { if } \beta \neq 1 \text { and } 0 \leq n<N_g \\ \frac{N_g-n}{N_g+1} & \text { if } \beta=1 \text { and } 0 \leq n<N_g \\ 1 & \text { if } n<0 \\ 0 & \text { if } n \geq N_g .
    \end{cases}
\end{align}
As a result, assuming the official blockchain extends $N$ blocks and the malicious $n_Y$, the attacker commences the race trailing by $(N-n_Y)$ blocks. In this case, the probability of a successful alternative history attack can be expressed~as
\begin{align}
    \begin{split}
        & \mathrm{S}\left(N, \beta, N_g\right) \\
        & =\sum_{n_Y=0}^{\infty} \operatorname{Pr}\left\{\text{Win} \mid z=N-n_Y\right\} \operatorname{Pr}\left\{Y=n_Y ; N, \frac{1}{1+\beta}\right\} ,
        \label{Eq:Sum_attack_1}
    \end{split}
\end{align}
or equivalently
\begin{align}
    \begin{split}
        & \mathrm{S}\left(N, \beta, N_g\right) \\
        & =\sum_{n_Y=0}^{\infty}\binom{n_Y+N-1}{n_Y}\left(\frac{1}{1+\beta}\right)^N\left(\frac{\beta}{1+\beta}\right)^{n_Y} P_{N-n_Y} .
        \label{Eq:Sum_attack}
    \end{split}
\end{align}
Finally, by exploiting the identity
\begin{align}
    \sum_{n=0}^{\infty}\binom{n+N-1}{n}\left(\frac{1}{1+\beta}\right)^Nd\left(\frac{\beta}{1+\beta}\right)^n=1 ,
\end{align}
\eqref{Eq:Sum_attack} can be rewritten as~\eqref{Eq:P_attack1}. We can conclude that the success of an attack it depends on a plethora of parameters such as the hash power of the attacker, the $N_g$ threshold, the mining rate $R_m$ of the official blockchain, as well as the official's confirmation number $N$. A higher $N$ value, while can be slowing down the speed of the blockchain and increase the total latency of the model, at the same time it increases the security of it, by challenging the malicious chain to keep the pace with the official. Additionally, by adopting the longest-chain rule in the official, priority is given to the value of proof of work as this approach is widely recognized for maintaining data integrity and making malicious attacks more difficult as they require greater computational power (hash power) by the attackers~\cite{Nakamoto2009}. \color{black}

\begin{figure*}
    \begin{align}
        \mathrm{S}\left(N, \beta, N_g\right)= \begin{cases}1-\sum_{n=0}^N\binom{n+N-1}{n}\left(\frac{1}{1+\beta}\right)^N\left(\frac{\beta}{1+\beta}\right)^n\left(\frac{1-\beta^{N-n+1}}{1-\beta^{N_g+1}}\right) & \text { if } \beta \neq 1 \\ 1-\sum_{n=0}^N \frac{1}{2^{N+n}}\binom{n+N-1}{n}\left(\frac{N-n+1}{N_g+1}\right) & \text { if } \beta=1\end{cases}
        \label{Eq:P_attack1}
    \end{align}
    \hrulefill 
\end{figure*}

%\begin{figure*}
%    \begin{align}
%       S(N, \beta)= \begin{cases}1-\sum_{n=0}^N\left(\begin{array}{c}
%        n+N-1 \\
%        n
%        \end{array}\right)\left(\frac{1}{1+\beta}\right)^N\left(\frac{\beta}%{1+\beta}\right)^n\left(1-\beta^{N-n+1}\right) & \text { if } \beta<1 \\
%        1 & \text { if } \beta \geq 1\end{cases}
%        \label{Eq:P_attack}
%    \end{align}
%\end{figure*}

\section{NUMERICAL RESULTS} \label{S:numerical_results} 
In this section, we present numerical results obtained from the proposed B-RAN model alongside interesting discussions that assess its performance and highlight valuable design guidelines. \color{black}It is worth noting that our simulation scenarios required substantial process power to manage the complex calculations. The following plots have produced results based on multiple simulations and events ($10^6$) to ensure the statistical validity and reliability of the results. To achieve this, we utilized a high-performance computing environment in order to execute our multiple scenarios properly and extract our results. Below is listed the hardware that we have used to fulfill our computational needs: CPU: AMD Ryzen 5 7600X 6-Core Processor 4.70 GHz, GPU: NVIDIA RTX 4070 Ti, RAM: 32 GB, OS: Linux Ubuntu 24.01. \color{black}By utilizing multiple types of carefully constructed figures, we provide a visual representation of the latency of the model and contrast it with traditional models. These results highlight how closely the BRAN model captures latency while also offering insights about its flexibility and scalability in realistic scenarios with different topologies. Furthermore, we investigate the security features of the proposed B-RAN model and demonstrate its robustness against possible attacks. The presented results serve as a prism that highlights the complexities of B-RAN and helps to derive definitive conclusions about its effectiveness and robustness in realistic usage scenarios.

\subsection{Latency}
\begin{figure}
    \centering
    \includegraphics[width=1\linewidth]{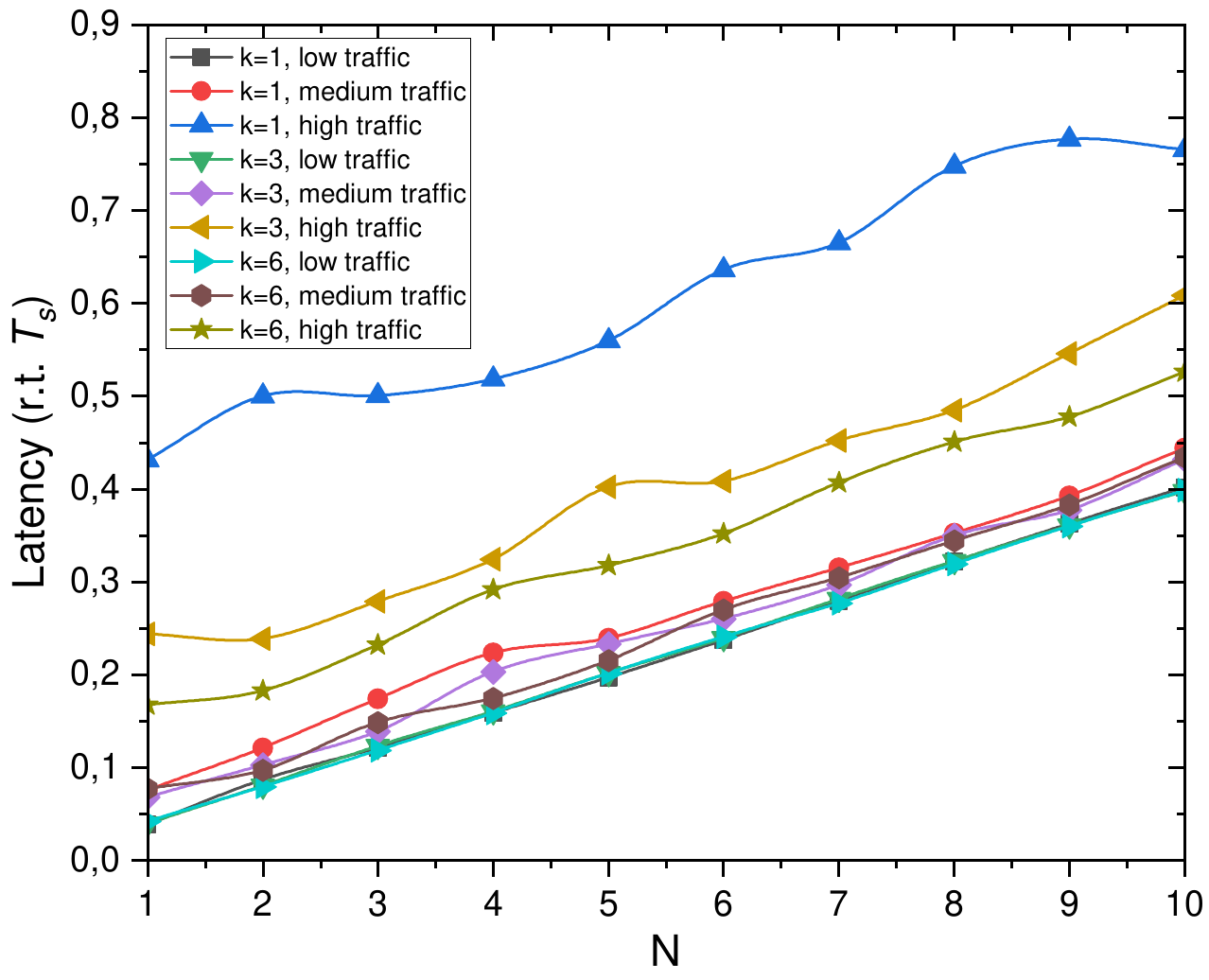}
    \caption{Latency vs $N$ for multiple $k$ and $\rho$ combinations.}
    \label{fig:latency_vs_N_3kModels}
\end{figure}
Fig.~\ref{fig:latency_vs_N_3kModels} presents the system's latency as a function of the number of confirmations, $N$, for different combinations of $k$ and $\rho$. For all plotted lines, we observe that, as the number of confirmations increases, the achievable latency also increases. Moreover, it becomes obvious that the scenarios with higher $k$ values exhibit better temporal performance. Specifically, for the high traffic regime, the model with $k=6$ has the lowest latency, while the conventional model with $k = 1$ the worst. Finally, it is worth noting that in the low intensity case all scenarios achieve similar performance with no noticeable changes. This suggests that higher block capacity may appear insignificant when the system performs under reasonable or low traffic, but can have higher impact on latency in high traffic cases.

\begin{figure}
    \centering
    \includegraphics[width=1\linewidth]{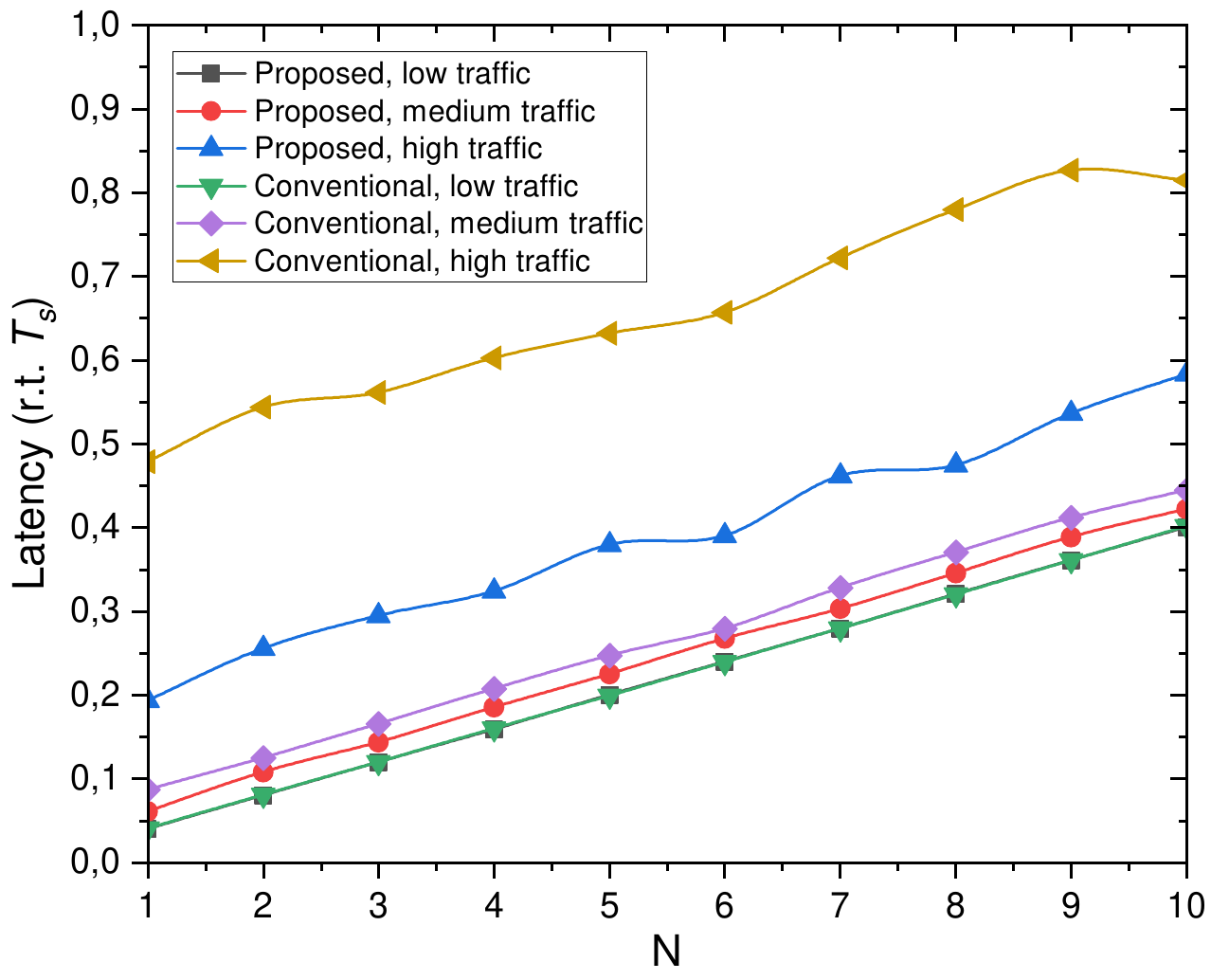}
    \caption{Latency vs $N$ in single blockchain}
    \label{fig:latency_vs_N}
\end{figure}
Fig.~\ref{fig:latency_vs_N} presents a comparison between the conventional model and the proposed framework with regard to the achievable latency under different traffic intensity scenarios and different $N$ values. By observing this figure, it becomes evident that as the number of confirmations for mining a block increases, the latency increases as well. However, a deeper look reveals some significant differences between the proposed and the traditional models. Despite achieving similar performance in low-traffic scenarios, the traditional model is characterized by higher delay under medium and heavy traffic conditions. This highlights that the proposed framework is more capable of modeling the temporal performance of complex B-RAN systems under a plethora of traffic conditions. 

\begin{figure}
    \centering
    \includegraphics[width=1\linewidth]{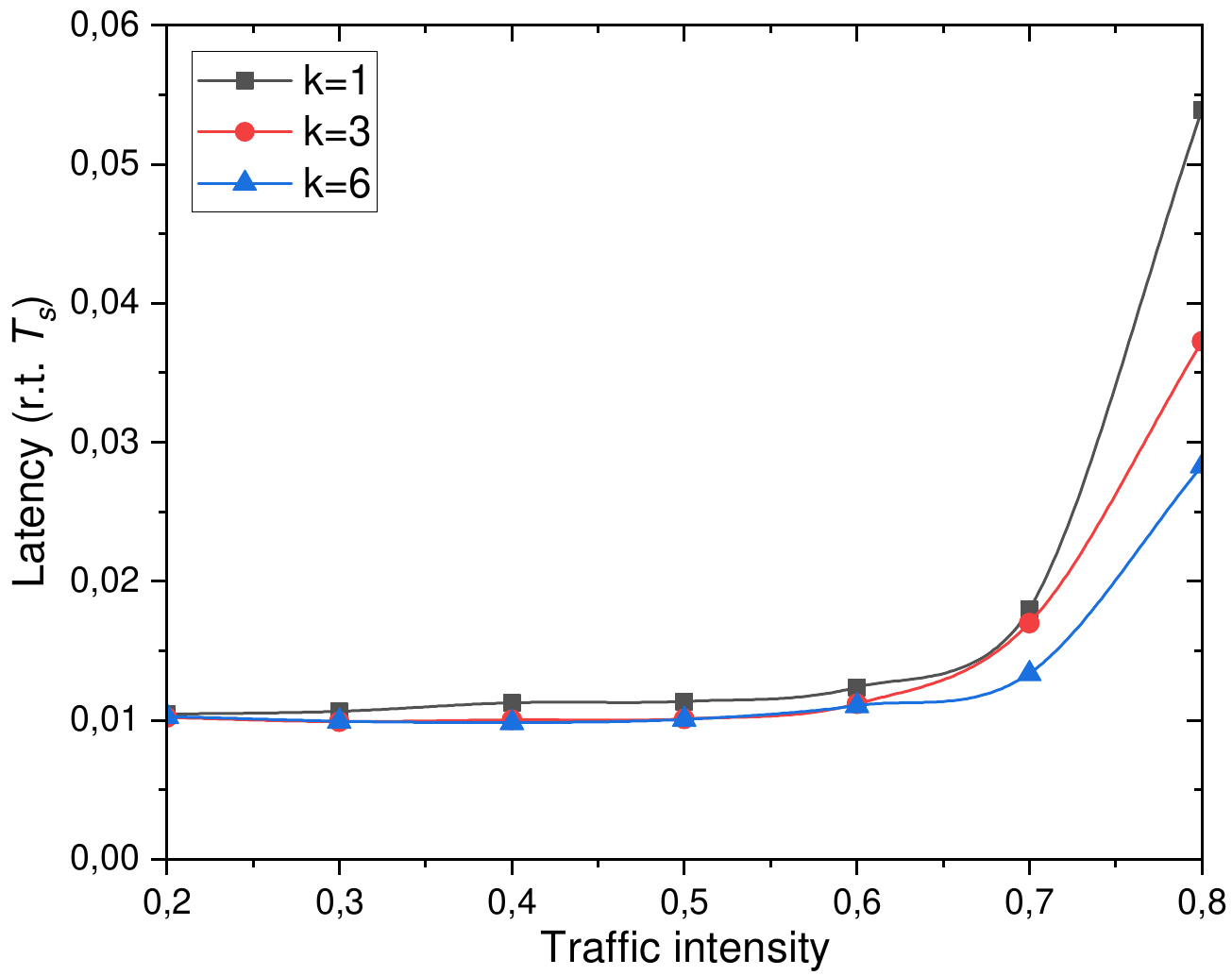}
    \caption{Latency vs traffic intensity for single blockchain for different $k$ values.}
    \label{fig:latency_vs_intensity_onlyprimary}
\end{figure}
Fig.~\ref{fig:latency_vs_intensity_onlyprimary} illustrates the latency as a function of the traffic intensity for different values of $k$ in the single blockchain scenario. From this figure, it becomes evident that the achievable latency for the low traffic regime achieves very close latency independent of the value of $k$. Moreover, until the threshold of $\rho = 0.5$, all cases exhibit the same behaviour. On the contrary, when traffic increases past this point, it becomes evident that the lines diverge from each other, with $k=1$ increasing significantly higher latency than the other two cases. For traffic intensity equal to $0.8$, we observe the biggest difference between the values. Specifically, the $k = 1$ line achieves the highest latency, while $k = 6$ has the lowest. This suggests that by increasing the number $k$ we achieve significantly lower latency especially in high traffic scenarios. In essence, as long as the intensity remains below a certain threshold, the amount of possible transactions per block has no major impact on the system's latency. As the intensity increases, larger values of $k$ allow the system to reach its true potential.

\begin{figure}
    \centering
    \includegraphics[width=1\linewidth]{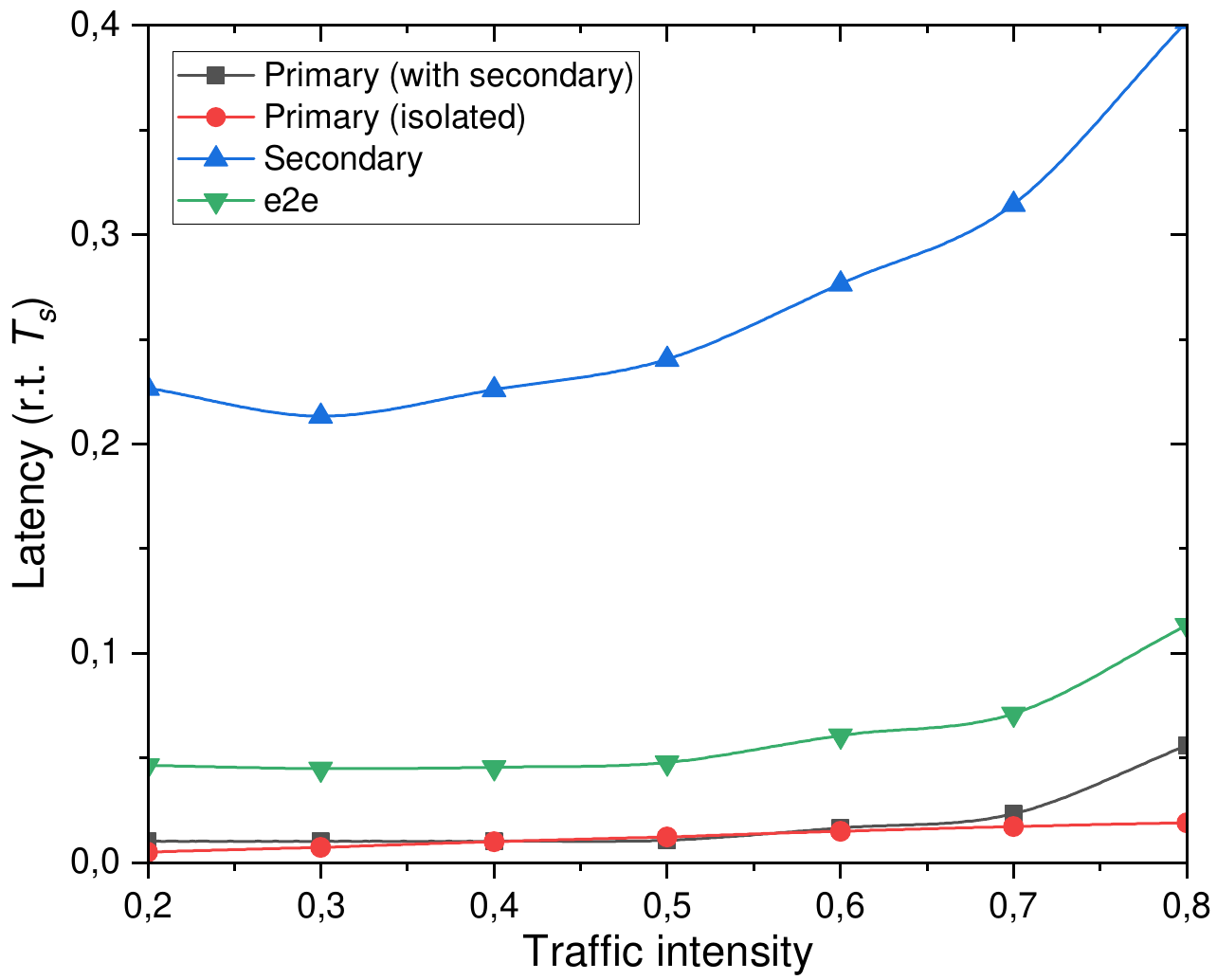}
    \caption{Latency vs traffic intensity of the HB-RAN.}
    \label{fig:latency_vs_r_first}
\end{figure}

A thorough study of traffic intensity impact on latency is presented in~Fig.~\ref{fig:latency_vs_r_first} that provides interesting insights on the average total delay of the system as well as the interactions of the primary and secondary blockchains. Specifically, the primary blockchain is analyzed both in conjunction with the secondary blockchain and on its own; thus, offering a deeper understanding of the performance degradation that is introduced by the hierarchical approach. Moreover, the various traffic intensity values were exclusively applied to the secondary blockchain in order to keep the primary blockchain's characteristics stable. This figure reveals a coherent trajectory, in which both blockchains and the average total delay exhibit a correlated increase in latency as the traffic intensity increases. Nonetheless, some subtle differences that characterize how primary and secondary blockchains behave differently from one another can be extracted. First and foremost, the secondary blockchain is characterized by significantly higher latency compared to the primary. \color{black}This is consequence of the fact that the primary chain has more computing capacity as it is the one connected with the BS, responsible for the direct connection between the users and the station, while the secondary has the responsibility to expand the network and connect the users with the primary. \color{black} In addition, as traffic intensity increases, the secondary blockchain experiences a steeper increase in delay, while the latency of the primary blockchain maintains relatively low latency. Moreover, the average total delay shows a larger dependency on primary blockchain performance, with a trajectory that is closer to the performance of the primary blockchain. Also, the plot highlights the efficiency of an isolated primary blockchain, devoid of secondary blockchain influence. In conclusion, the plot clarifies how variations in network traffic can affect temporal dynamics and the system's overall performance, offering important insights into the complex relationship between the primary and secondary blockchains.

\begin{figure*}
    \centering
    \begin{subfigure}{0.32\linewidth}
        \centering
        \includegraphics[width=1\linewidth]{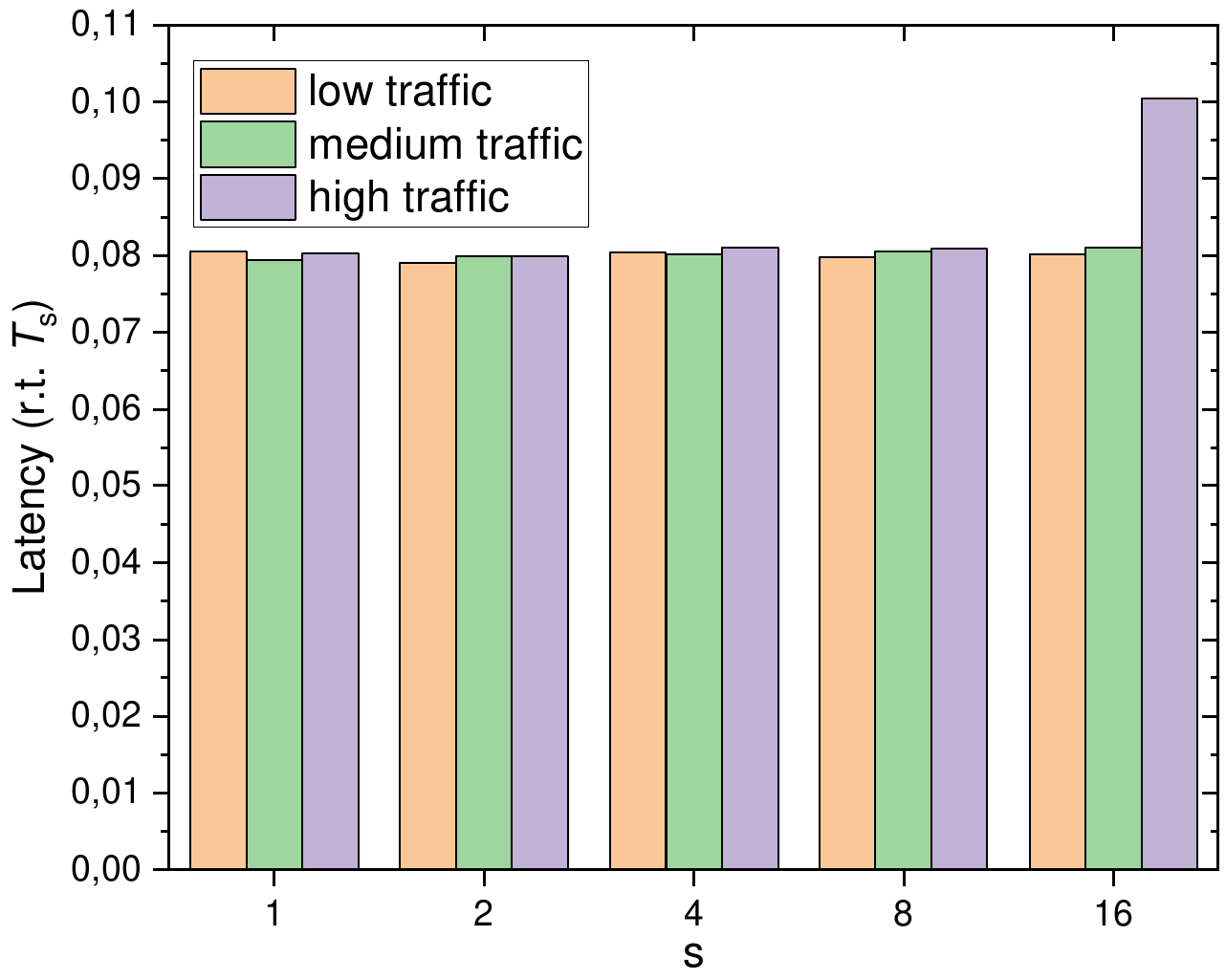}
        \caption{Latency (primary blockchain) for $k = 1$.}
        \label{fig:Primary_k1}
    \end{subfigure}
    \begin{subfigure}{0.32\linewidth}
        \centering
        \includegraphics[width=1\linewidth]{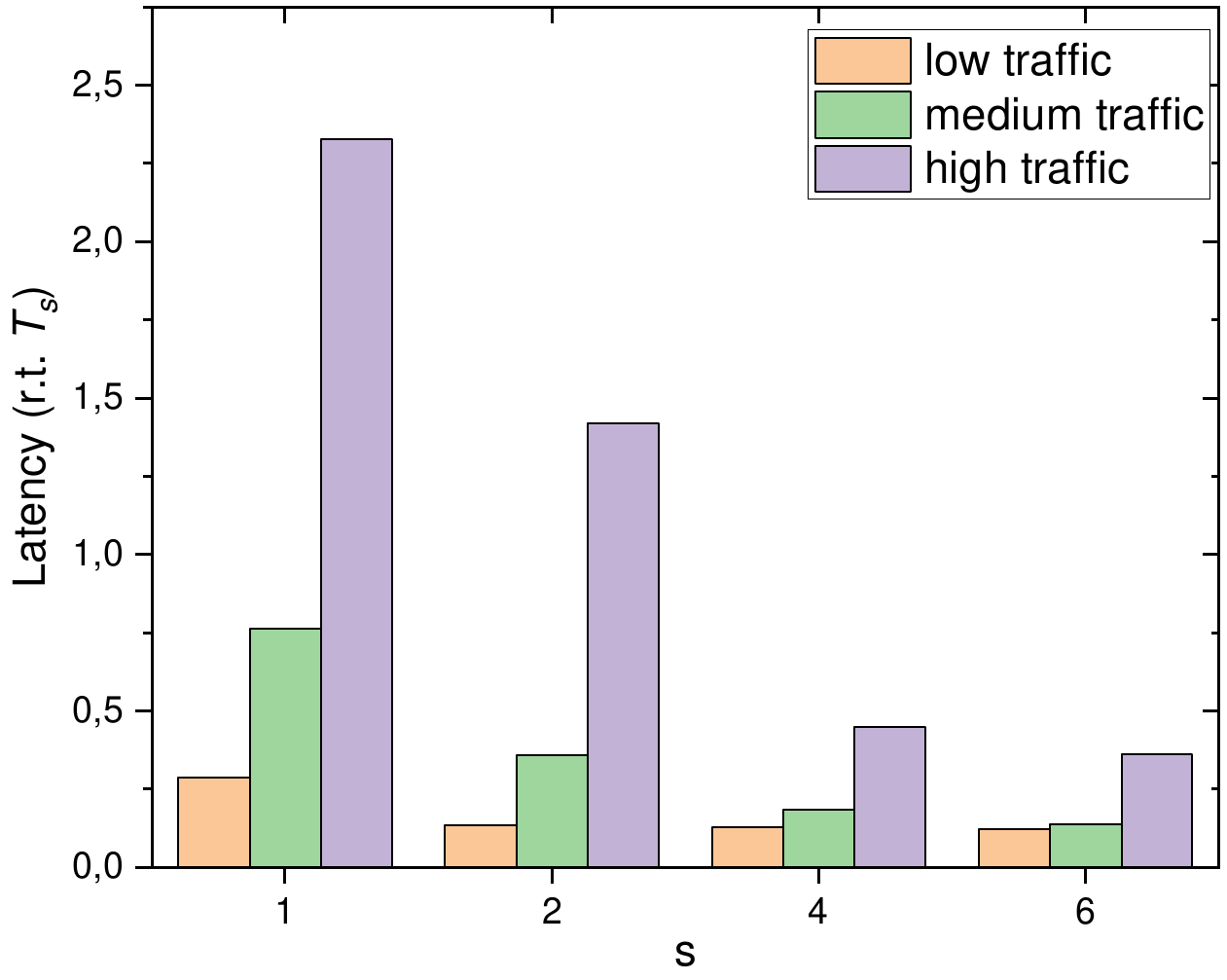}
        \caption{Latency (secondary blockchain) for $k = 1$.}
        \label{fig:latency_k1}
    \end{subfigure}
    \begin{subfigure}{0.32\linewidth}
        \centering
        \includegraphics[width=1\linewidth]{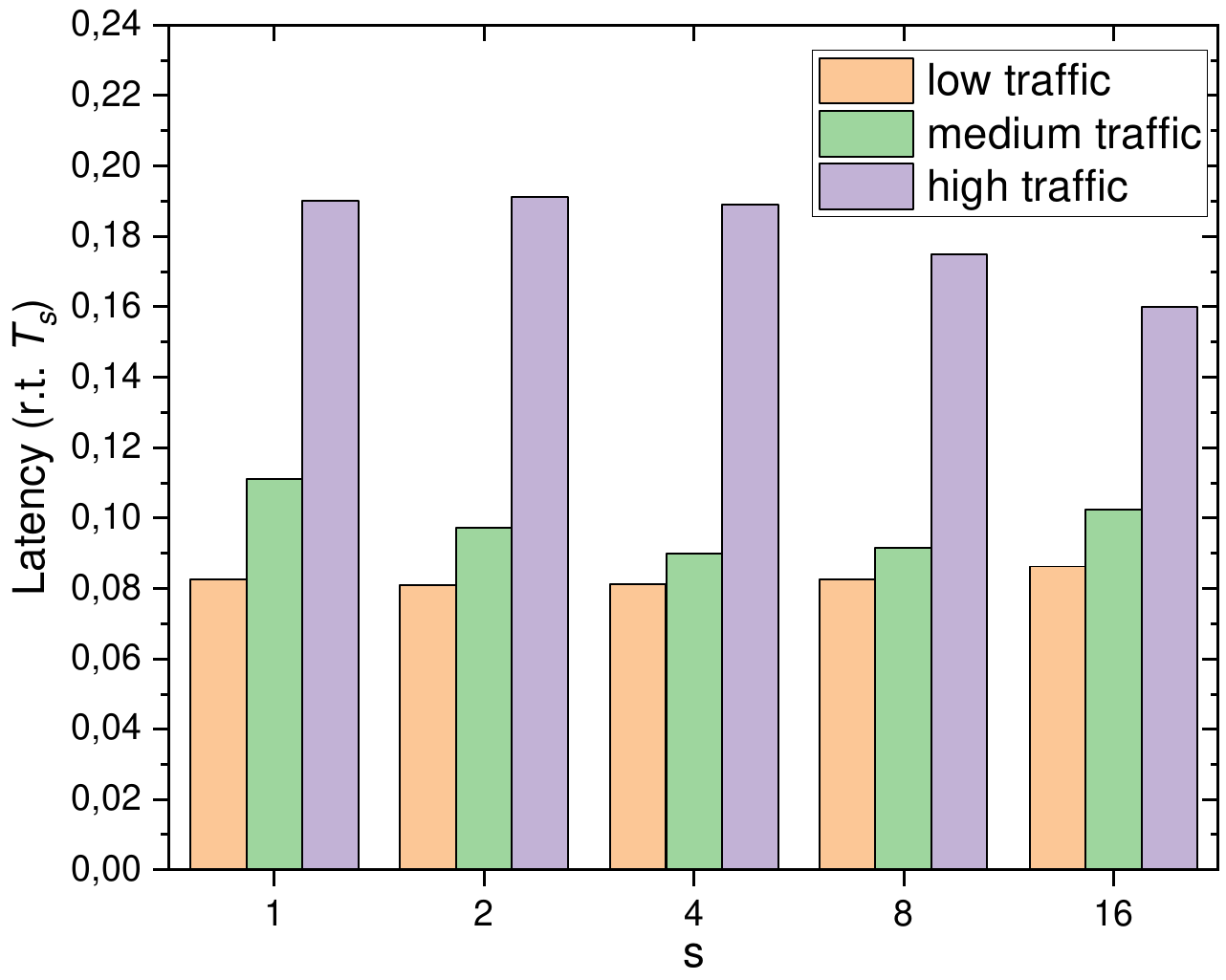}
        \caption{e2e latency for $k = 1$.}
        \label{fig:e2e_k1}
    \end{subfigure}
    \begin{subfigure}{0.32\linewidth}
        \centering
        \includegraphics[width=1\linewidth]{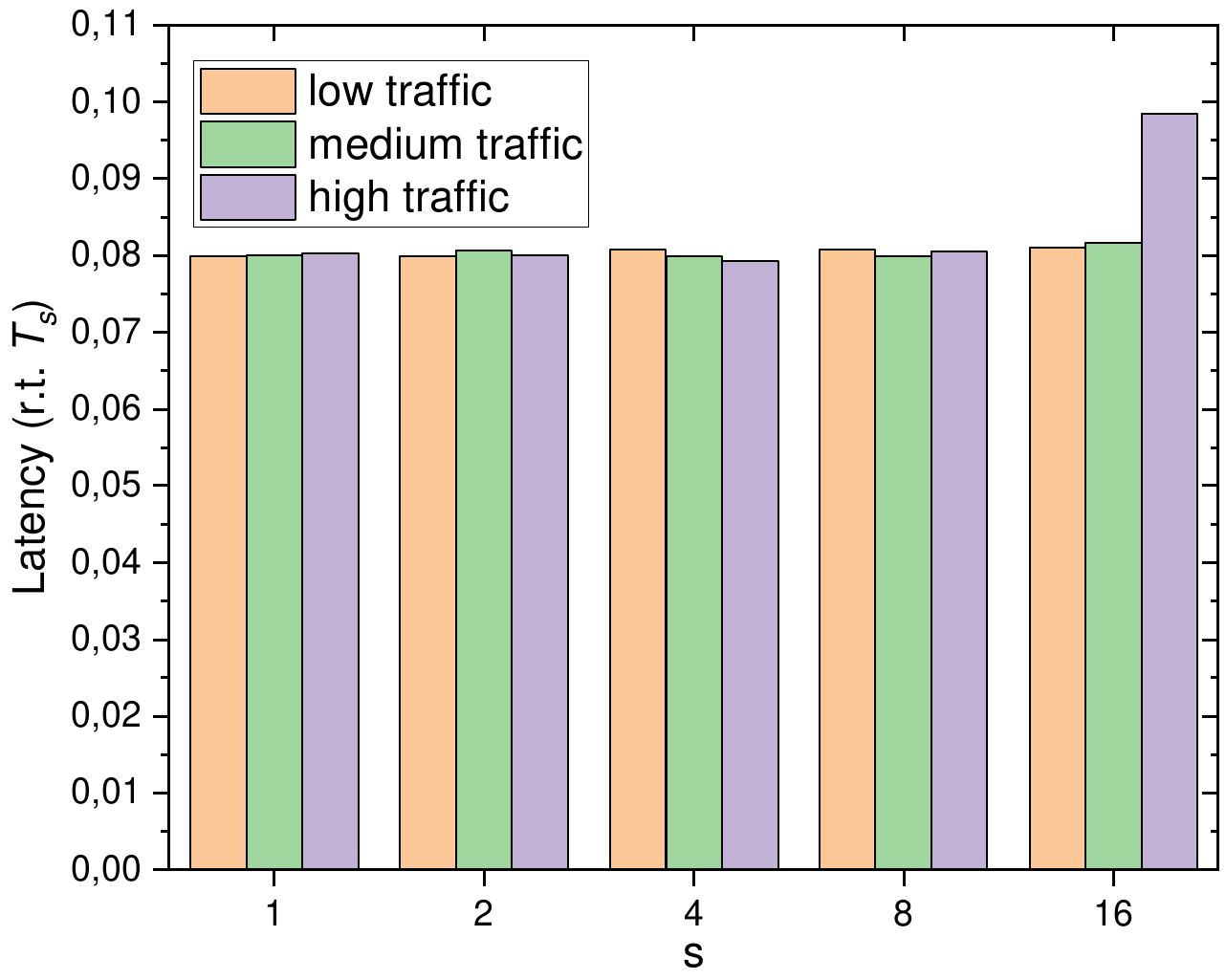}
        \caption{Latency (primary blockchain) for $k = 3$.}
        \label{fig:Primary_k3}
    \end{subfigure}
    \begin{subfigure}{0.32\linewidth}
        \centering
        \includegraphics[width=1\linewidth]{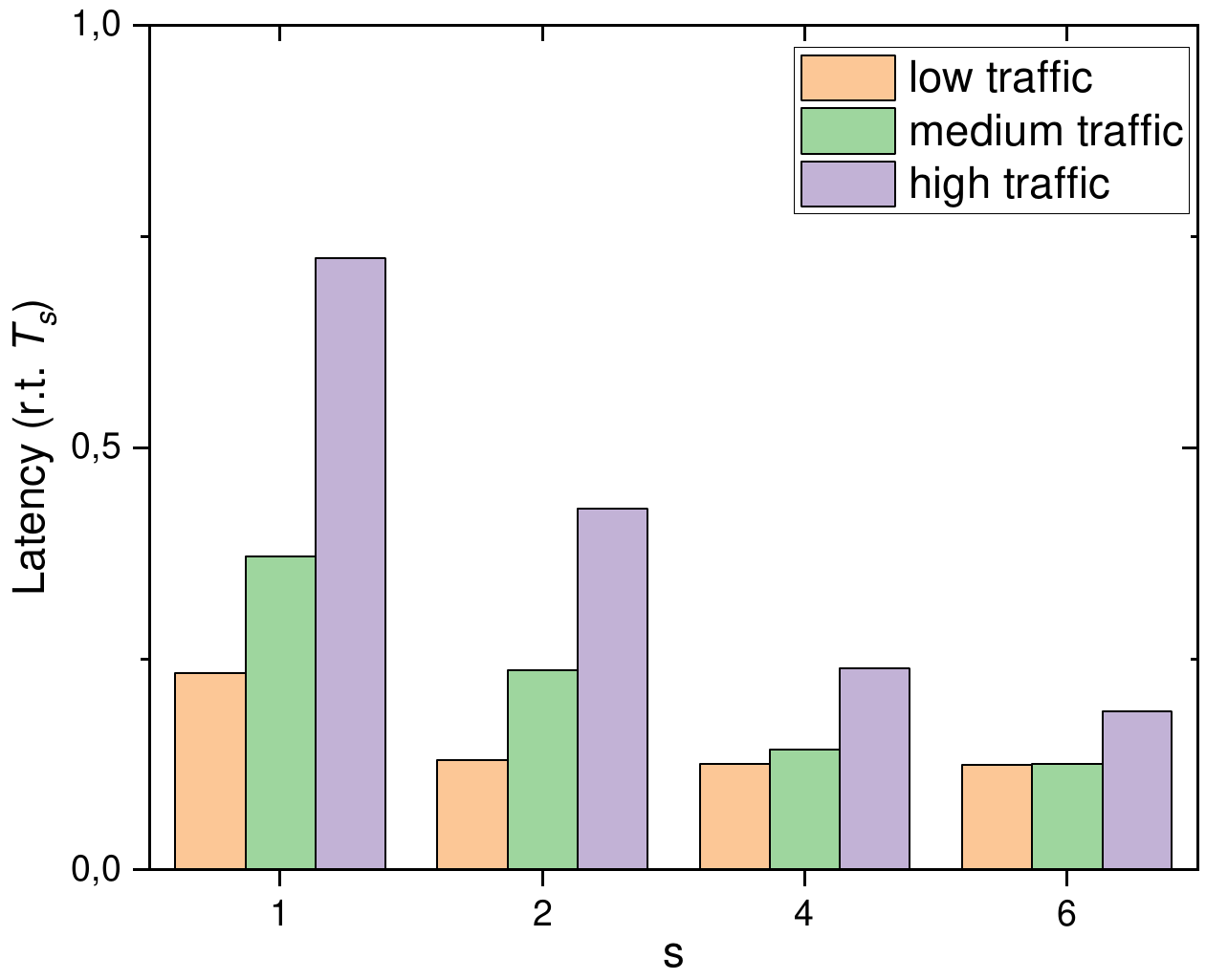}
        \caption{Latency (secondary blockchain) for $k = 3$.}
        \label{fig:latency_k3}
    \end{subfigure}
    \begin{subfigure}{0.32\linewidth}
        \centering
        \includegraphics[width=1\linewidth]{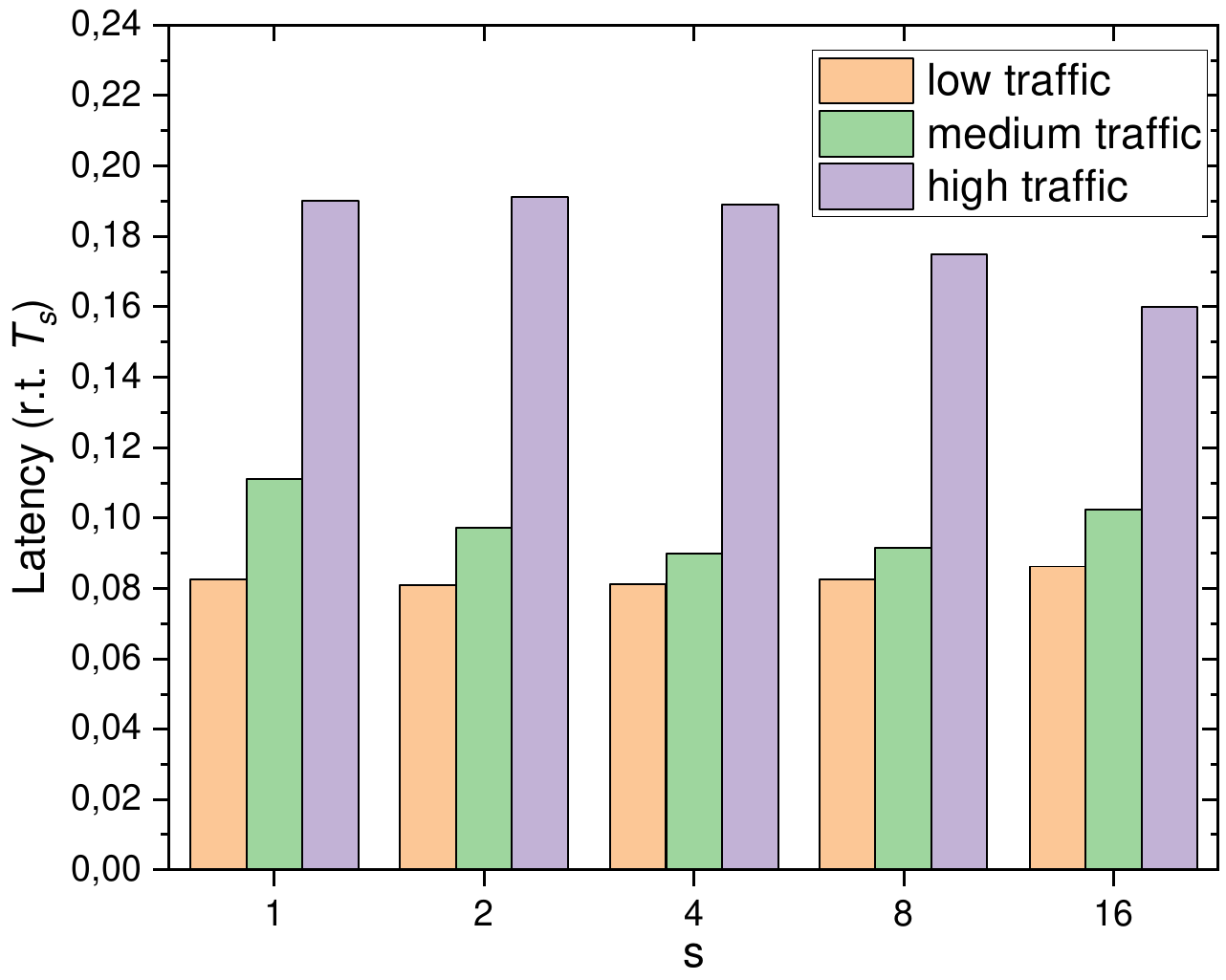}
        \caption{e2e latency for $k = 3$.}
        \label{fig:sub2}
    \end{subfigure}
    \caption{Primary, secondary, and e2e latency vs $S$ for various combinations of $k$ and $\rho$.}
    \label{fig:Mutli_graphs}
\end{figure*}
Fig.~\ref{fig:Mutli_graphs} presents the latency achieved by the primary and the secondary blockchains as well as the e2e system as a function of the number of concurrent users that can be served by a BS, $s$. In more detail, subfigures (a)-(c) assume that each block of the secondary blockchain can contain a maximum of $k = 1$ \color{black}transactions, \color{black}while the rest assume $k = 3$. By comparing the achievable latency of the primary blockchain as depicted in subfigures (a) and (d), it becomes evident that the latency is not affected by variations in the secondary blockchain's maximum capacity or traffic intensity. It is interesting to point out that for $s$ values higher than $16$, the latency increases significantly. This phenomenon is caused by the congestion brought on by the high traffic of the secondary blockchain that affects the primary one. Furthermore, the latency performance of the secondary blockchain is illustrated in subfigures (b) and (e), which showcase how greatly can $k$, $\rho$, and $s$ affect the performance of the secondary blockchain. Specifically, high traffic values caused increased latency, while low $k$ and $s$ values can restrict the system's performance to a great extend. This highlights the importance of appropriately selecting the various degrees of freedom when designing B-RAN systems. Finally, subfigures (c) and (f) present the e2e latency of the system, which is shown to be significantly influenced by the number of contracts that can be included in a single block of the secondary blockchain. This emphasizes the equilibrium that is forged between traffic intensity and block capacity.

\begin{figure}
    \centering
    \includegraphics[width=1\linewidth]{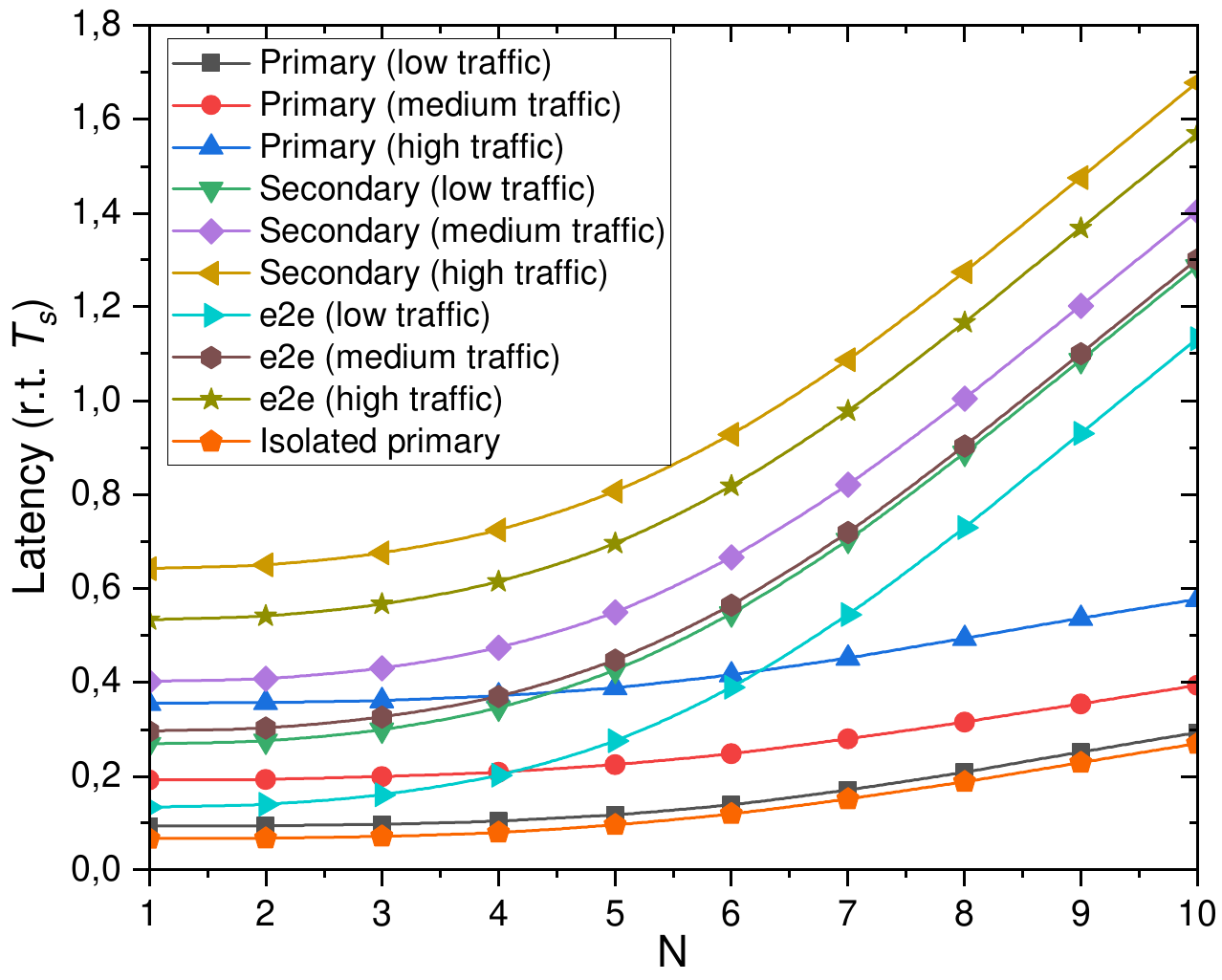}
    \caption{Latency vs $N$ of the hierarchical B-RAN with multiple intensities.}
    \label{fig:Latency vs N confirmations (multiple intensities)}
\end{figure}
An extensive examination of the impact of different $N$ confirmation numbers on latency is presented in~Fig.~\ref{fig:Latency vs N confirmations (multiple intensities)} for the primary and secondary blockchain. The plot shows the average total delay of the system along with the individual latency of the primary and secondary blockchains. Notably, the primary blockchain's behaviour remains constant throughout the plotted range of $N$, allowing for a focused examination of the impact of the secondary blockchain on the system. With three distinct traffic intensity values for the secondary blockchain — low, medium, and high — the plot demonstrates how various traffic scenarios influence the latency of the entire system. As anticipated, a higher traffic intensity correlates with increased latency across the board, while lower intensity results in lower latency. The average total delay of the system is particularly interesting, since it resembles the behaviour of the secondary blockchain. This discovery implies that secondary blockchain activity has a major impact on the overall performance of the system. Moreover, this figure provides an understanding of the extra latency that the secondary blockchain adds to the major blockchain, particularly when contrasting with a situation in which the primary blockchain is isolated. This comparison demonstrates how much secondary blockchain activity affects the parent blockchain's delay. Overall, the plot emphasizes the complex interplay between primary and secondary blockchains, showing how changes in the characteristics of the secondary blockchain can impact the temporal dynamics of the entire system.

\subsection{Security}
This section sheds light on the security aspects of the proposed B-RAN model by providing numerical results that collected based on the modelling the alternate history attack that was presented in~Section~\ref{S:Attack}. The demonstrated results focus on the interaction among various degrees of freedom of B-RAN and provide interesting discussions on the its adaptability to various configurations. This analysis is crucial, as attackers can target either the primary or secondary blockchain of the proposed framework. Moreover, the achieved security performance is compared to other works. This allows us to capture the dynamic nature of security challenges within B-RAN and provide design guidelines that ensure robust protection against potential attacks.

\begin{figure}
    \centering
    \includegraphics[width=1\linewidth]{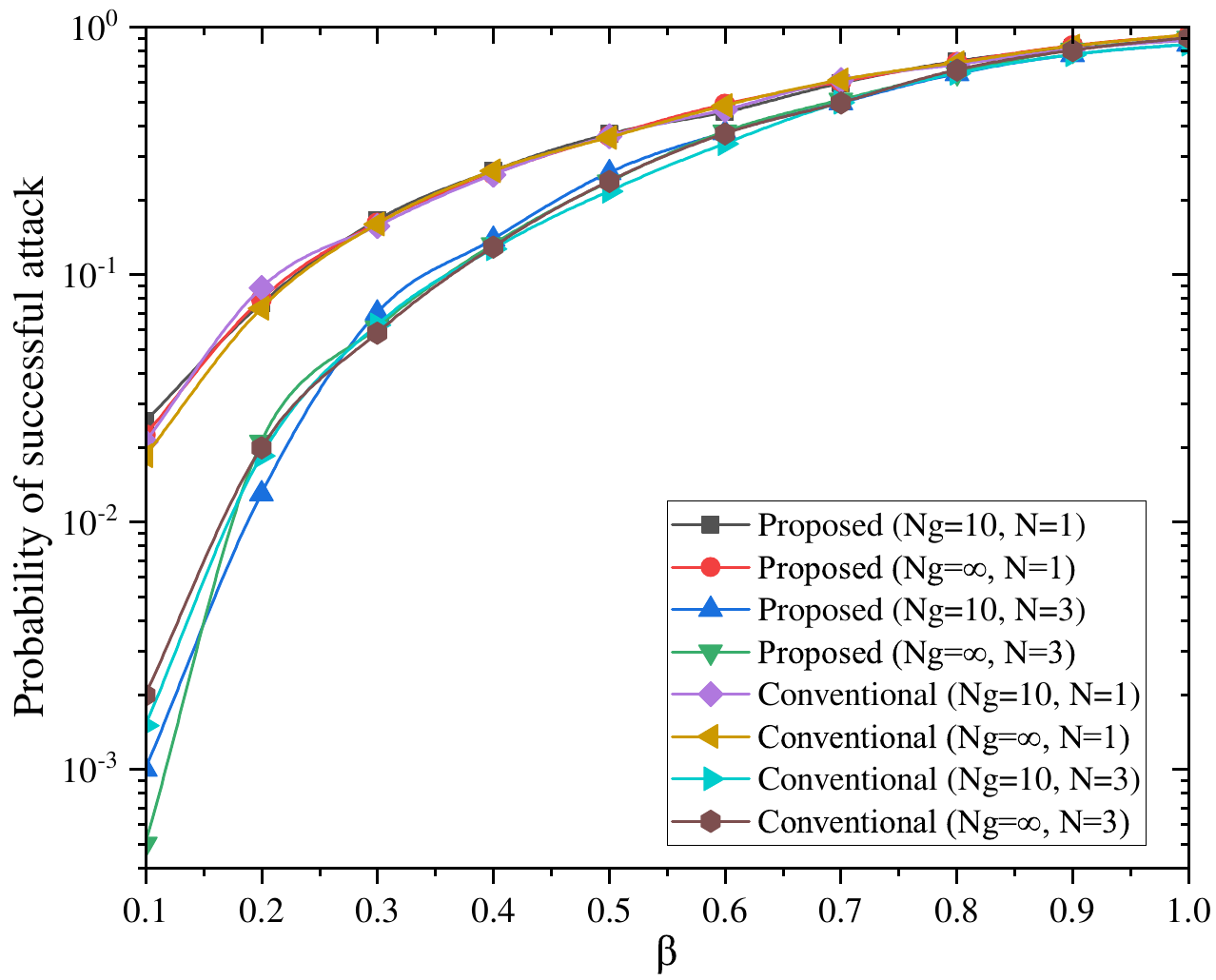}
    \caption{Probability of successful attack vs the attacker's mining power for multiple combinations of $N_g$ and $N$.}
    \label{Fig:attack}
\end{figure}
Fig.~\ref{Fig:attack} presents the probability of successful attack as a function of the rate between the hash power of the official and malicious blockchains. Different attack strategies, $N_g$, of the attacker and various numbers of confirmations, $N$, are taken into consideration in the analysis. All of the scenarios include both the proposed and conventional B-RAN modelling approaches. It immediately becomes evident that the two modelling approaches provide similar results, which validates the validity of the proposed framework. As expected, the probability of successful attacks increases as the $\beta$ values increase, while it approaches $100\%$ when the malicious and official blockchains have comparable mining power. This is the case for both $N = 1$ and $N = 3$ configurations, with the latter exhibiting better security performance of $2\times10^-3$ for low $\beta$ values. This observation indicates a consistent behavioral pattern for both configurations across various attack scenarios, regardless of variations in $N_g$ values. The convergence of the two configurations suggests a shared vulnerability for B-RAN systems that is introduced due to the existence of blockchain. Finally, it is important to highlight that the official blockchain is characterized by robust computational capabilities that cannot be easily matched by the malicious one. 

\begin{figure}
    \centering
    \includegraphics[width=1\linewidth]{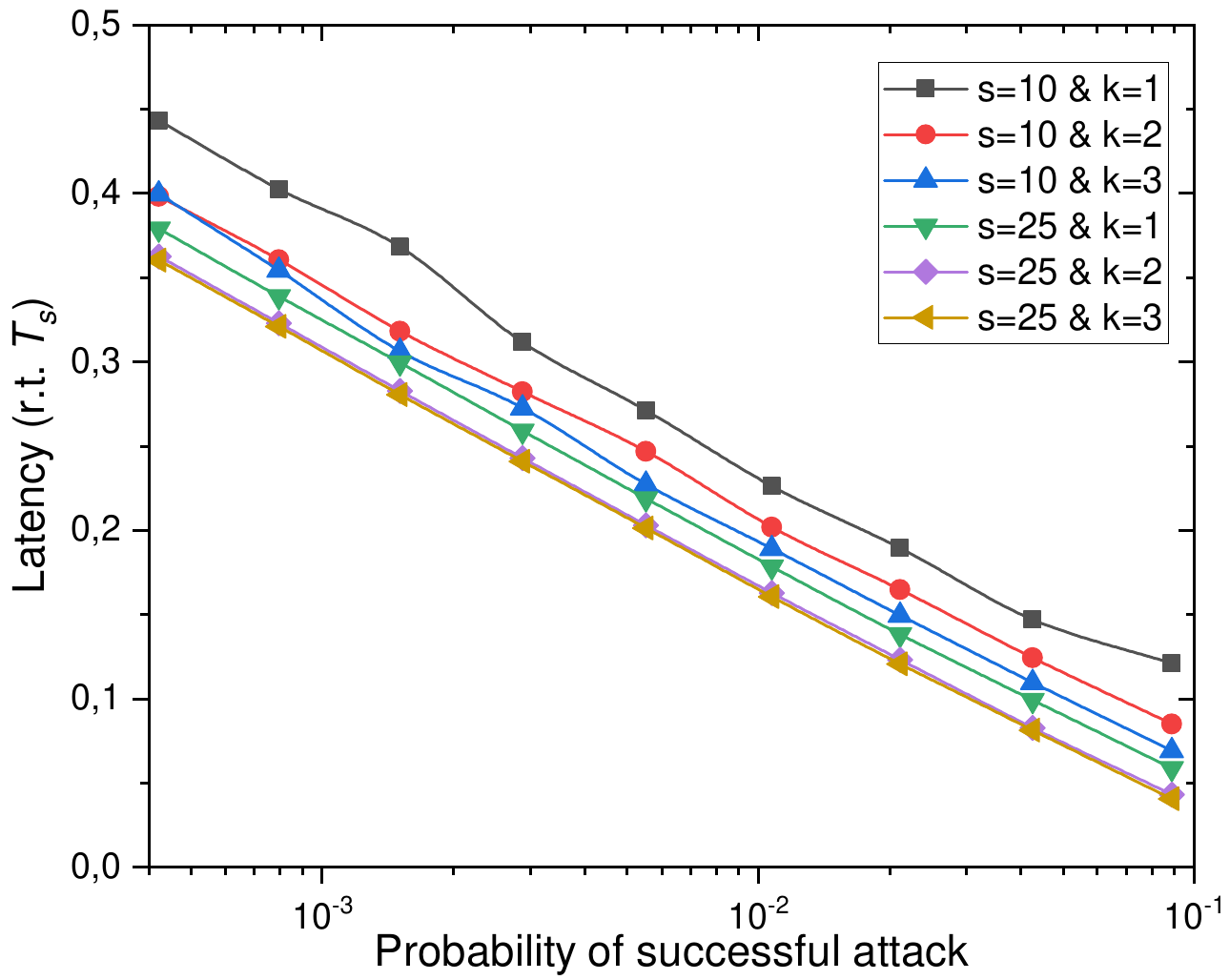}
    \caption{Security vs latency for different $s$ and $k$ configurations.}
    \label{Fig:securityVSlatency}
\end{figure}
Fig.~\ref{Fig:securityVSlatency} illustrates the interactions between latency and security for various different configurations of B-RAN. Specifically, six combinations are drawn for $s = 10$ or $25$ and $k$ values equal to $1$, $2$, or $3$. By observing any of the plotted configurations, it is evident that as security increases the latency increases as well. This highlights an equilibrium between security and latency when designing B-RAN systems. \color{black}In systems where security plays a significant role, a trade-off with temporal performance is expected, and vice versa. \color{black}Although at first glance different configurations appear to have similar behaviour, a closer inspection uncovers significant differences. Specifically, the black line that  is characterized by $k = 1$ and $s = 10$ indicates the worst performance in both security and latency, whereas the yellow line of $k = 3$ and $s = 25$ is the fastest and the most resilience to attacks. Furthermore, the models with higher $s$ values demonstrate a notable improvement in security when compared to their low-$s$ counterparts, which highlights the important role of $s$. Moreover, a closer investigation of the $s = 25$ configurations reveals a medium variation between $k = 1$ and $k = 2$, but only a minor difference between $k = 2$ and $k = 3$. This indicates that after a certain point increasing the $k$ value does not result in improved performance. Overall, this figure underlines the significance of appropriately selecting the design parameters of B-RAN to achieve the intended latency without sacrificing security.

\section{CONCLUSIONS \& FUTURE DIRECTIONS} \label{S:conclusions}
\color{black} This paper proposes HB-RAN, a hierarchical blockchain-based radio access network that meets the key challenges in wireless connectivity by integrating blockchain with MEC and AI. In the HB-RAN architecture, the role of participants in the network is redefined by enabling service consumers to also act as providers, leading to collaboration for efficient use of resources. The incorporation of blockchain into the system ensures secure and trustworthy communication, overcoming the limitations of traditional networks. 

The Markov chain theoretical model that was introduced provides valuable insights into the achievable performance of B-RAN; thus, enabling the accurate evaluation of both latency and security in three real-world usage scenarios, namely fronthaul network of fixed topology, advanced coverage expansion, and advanced connectivity of mobile nodes. Although blockchain systems contain complex interactions between nodes, user equipment, and the network, the proposed framework is a flexible and versatile method with many degrees of freedom that constitute it capable of being applied to many different blockchain deployments. This novelty research is the first to explore the extension of a BRAN model based on Markovian chains with this degree of parameterization. Numerical results were produced that validate the theoretical modeling through simulations and highlight that the architecture not only mitigates security risks but also supports seamless connectivity. 

The real-world implications of HB-RAN are significant, especially in the context of autonomous transportation, smart cities, and remote communication. By supporting advanced use cases such as V2V communication, rural connectivity, and AI-driven services, it can ensure further scalability, energy efficiency, and system reliability. Enabling innovation in the most important areas, HB-RAN spans a wide range of applications-from fixed fronthaul to coverage extension and mobile connectivity, making it a key enabler for next-generation 6G networks. Our future work will focus on real-world implementation and testing of B-RAN to further validate the proposed theoretical framework.\color{black}

%\balance
\bibliographystyle{IEEEtran}
\bibliography{refs}

\begin{IEEEbiography}
    [{\includegraphics[width=1in,height=1.25in,clip,keepaspectratio]{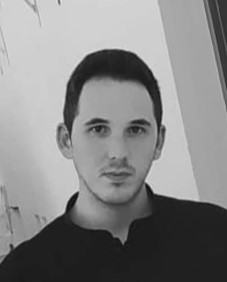}}]
    {Vasileios Kouvakis } was born in Thessaloniki, Greece in 1995. He received his diploma in Informatics \& Telecommunications Engineering (ICTE) (5 years) from University of Western Macedonia (UOWM) in 2021. From May 2023 until now, he is working at InnoCube as an Artificial Intelligence (AI) Engineer \& Research associate. His interests include AI, Machine Learning (ML), as well as Communication \& Networks.
\end{IEEEbiography}

\begin{IEEEbiography}
    [{\includegraphics[width=1in,height=1.25in,clip,keepaspectratio]{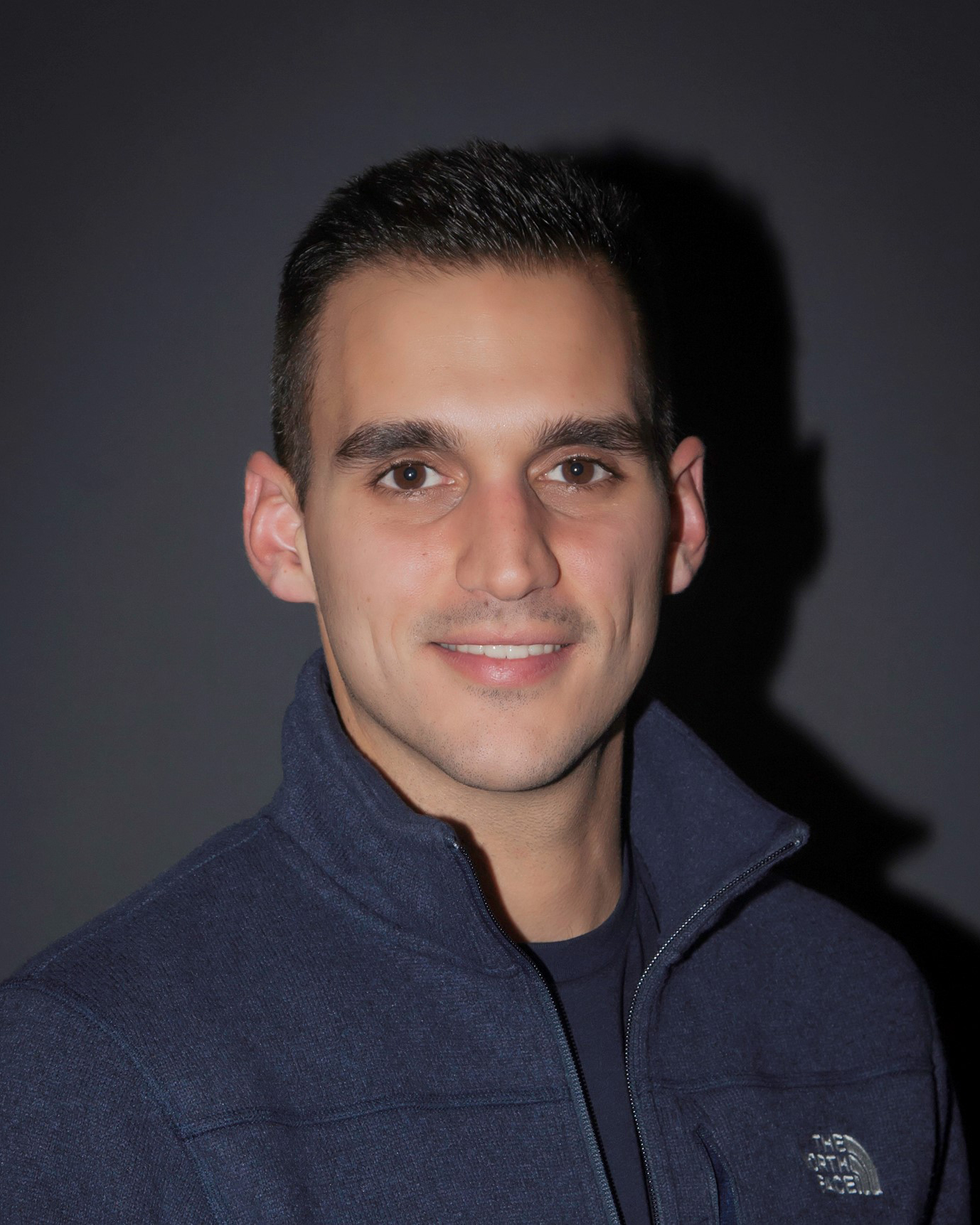}}]
    {Stylianos E. Trevlakis } (Member, IEEE) was born in Thessaloniki, Greece in 1991. He received the Electrical and Computer Engineering (ECE) diploma (5 year) from the Aristotle University of Thessaloniki (AUTh) in 2016. Afterwards, Dr. Trevlakis served in the Hellenic Army in for nine months in the Research Office as well as at the Office of Research and Informatics of the School of Management and Officers. During 2017, he joined the Information Technologies Institute, while from October 2017 until April 2022, he was part of WCIP as a PhD candidate in AUTh. During the same period, he was a teaching assistant at the department of ECE of AUTh. From April 2022 until now, Dr. Trevlakis is working at InnoCube as a postdoctoral researcher with focus on state-of-the-art research in conventional \& AI-enabled Wireless Communication Systems.

    Dr. Trevlakis' research interests lie in the area of Wireless Communications, with emphasis on conventional \& AI-enabled Wireless Communication Systems, as well as Communications \& Signal Processing for Biomedical Engineering.
\end{IEEEbiography}

\begin{IEEEbiography}
    [{\includegraphics[width=1in,height=1.25in,clip,keepaspectratio]{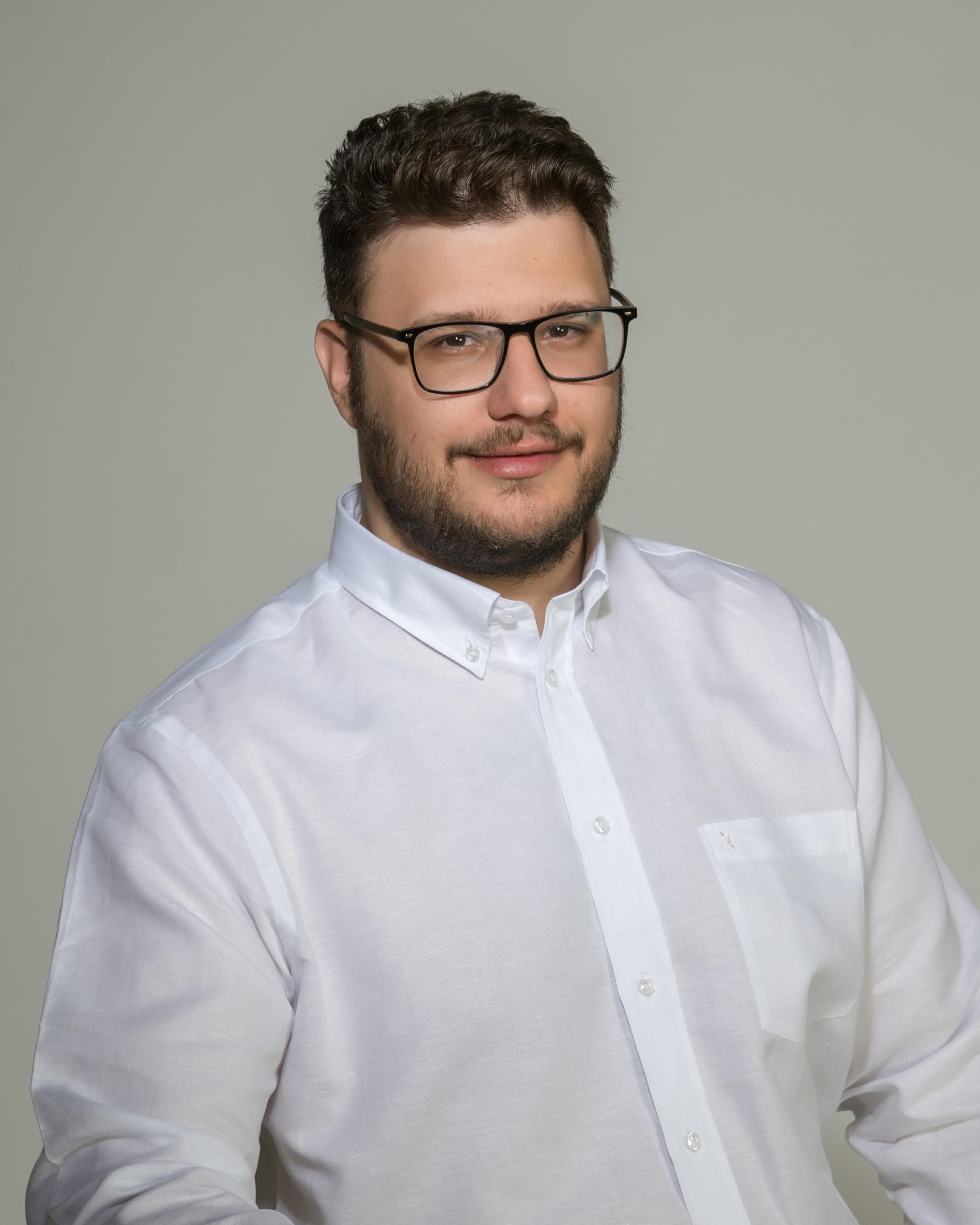}}]
    {Alexandros-Apostolos A. Boulogeorgos } (Senior Member, IEEE) received the Diploma degree in Electrical and Computer Engineering and the Ph.D. degree in wireless communications from the Aristotle University of Thessaloniki in 2012 and 2016, respectively. From 2022, he is an Assistant Professor at the Department Electrical and Computer Engineering of the University of Western Macedonia, Greece. Dr Boulogeorgos has (co-)authored more than 160 technical papers, which were published in scientific journals and presented at prestigious international conferences. Furthermore, he is the holder of 2 (1 national and 1 European) patents, while he has filled other 3 patents.  He is listed in ``World’s Top 2\% Scientists for the Year 2022'' and ``World’s Top 2\% Scientists for the Year 2023'' which is published by Stanford University and Elsevier.  His current research interests spans in the area of wireless communications and networks with emphasis in high frequency communications, intelligent communication systems with emphasis to semantic communications, optical wireless communications, and signal processing and communications for biomedical applications. 
    
    He has been involved as a member of organizational and technical program committees in several IEEE and non-IEEE conferences and served as a reviewer in various IEEE journals and conferences. He is an IEEE Senior Member and a Member of the Technical Chamber of Greece. He is currently an Editor for IEEE TRANSACTIONS ON WIRELESS COMMUNICATIONS, IEEE COMMUNICATIONS LETTERS, Frontier in Communications and Networks, and MDPI~Telecom.  Dr Boulogeorgos has participated as a guest editor in the organization of a number of special issues in IEEE and non-IEEE journals. 
\end{IEEEbiography}

\begin{IEEEbiography}
    [{\includegraphics[width=1in,height=1.22in,clip,keepaspectratio]{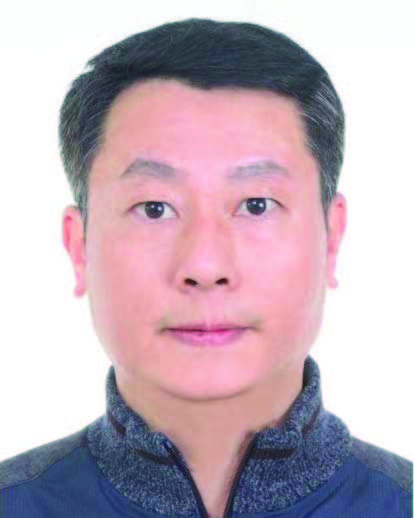}}]{Hongwu Liu }\, (Senior Member, IEEE) received the Ph.D. degree from Southwest Jiaotong University in 2008. From 2008 to 2014, he was with Nanchang Hangkong University. From 2010 to 2011, he was a Post-Doctoral Fellow with the Shanghai Institute of Microsystem and Information Technology, Chinese Academy of Science. From 2011 to 2013, he was a Research Fellow with the UWB Wireless Communications Research Center, Inha University, South Korea. Since 2014, he has been an Associate Professor with Shandong Jiaotong University. His research interests include MIMO signal processing, cognitive radios, cooperative communications, wireless secrecy communications, and AI-based wireless communications.
\end{IEEEbiography}

\begin{IEEEbiography}
[{\includegraphics[width=1in,height=1.25in,clip,keepaspectratio]{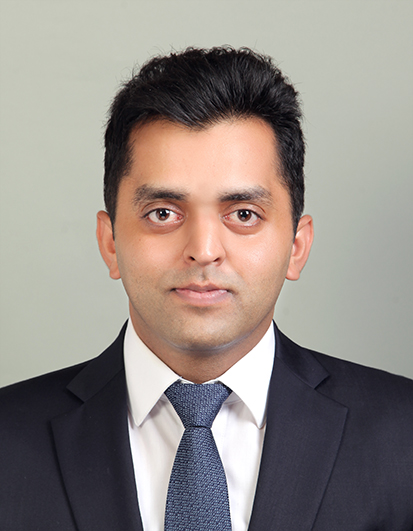}}]
   {Waqas Khalid } received the B.S. degree in Electronics Engineering from GIK Institute of Engineering Sciences and Technology, KPK, Pakistan, in 2011. He received M.S. and Ph.D. degrees in information and communication engineering from Inha University, Incheon, South Korea, and Yeungnam University, Gyeongsan, South Korea, in 2016, and 2019, respectively. He is currently working as a research professor at the Institute of Industrial Technology, Korea University, Sejong, South Korea. His areas of interest include physical layer modeling, signal processing, and emerging technologies for 5G networks, including reconfigurable intelligent surfaces, physical-layer security, NOMA, cognitive radio, UAVs, and IoTs. 
\end{IEEEbiography}

\begin{IEEEbiography}
[{\includegraphics[width=1in,height=1.25in,clip,keepaspectratio]{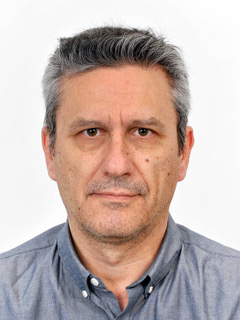}}]
    {Theodoros A. Tsiftsis } (S'02, M'04, SM'10) received the PhD degree in electrical engineering from the University of Patras, Greece, in 2006. He is a professor with the Department of Informatics \& Telecommunications, University of Thessaly, Greece, and also an honorary professor with Shandong Jiaotong University, Jinan City, China. His research interests fall into the broad areas of communication theory and wireless communications, with an emphasis on wireless communications theory, reconfigurable intelligent surfaces, optical wireless communications, and physical layer security.
    
    Dr. Tsiftsis served on the Editorial Boards of the \textsc{IEEE Transactions on Communications}, \textsc{IEEE Transactions on Vehicular Technology}, \textsc{IEEE Communications Letters}, and \textsc{IEEE Transactions on Mobile Computing}. He is currently an Associate Editor of the \textsc{IEEE Transactions on Wireless Communications}, and Specialty Chief Editor for Networks and Communications of Frontiers in Computer Science. Prof. Tsiftsis was appointed as an IEEE Vehicular Technology Society Distinguished Lecturer for two terms (2018–2022) and was recently appointed as an IEEE Communications Society Distinguished Lecturer (2024–2025). 
\end{IEEEbiography}

\begin{IEEEbiography}
 [{\includegraphics[width=1in,height=1.25in,clip,keepaspectratio]{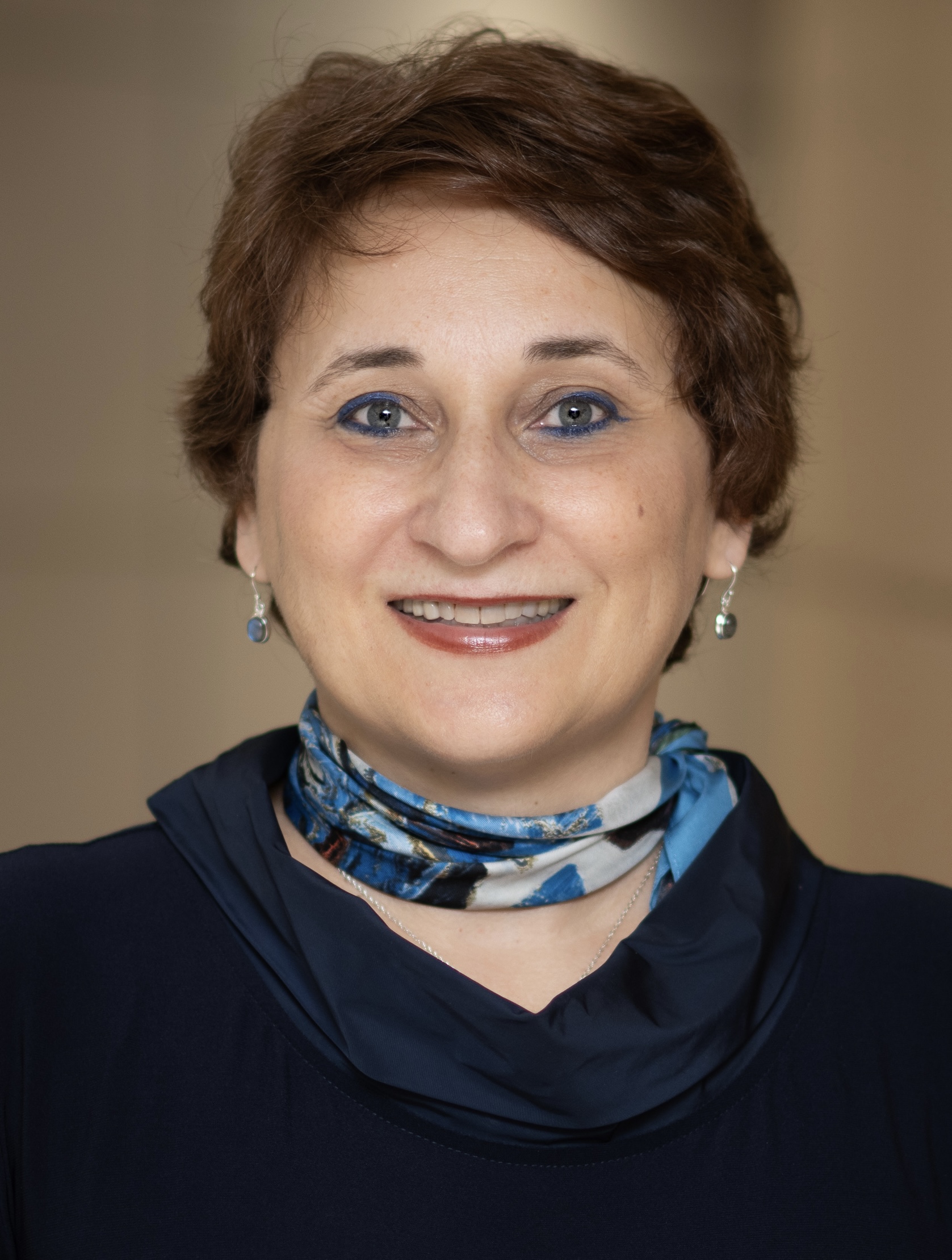}}]
    {Octavia A. Dobre }  (Fellow, IEEE) is a Professor and Tier-1 Canada Research Chair with Memorial University, Canada. She was a Visiting Professor with Massachusetts Institute of Technology, USA and Université de Bretagne Occidentale, France.Her research interests encompass wireless communication and networking technologies, as well as optical and underwater communications. She has (co-)authored over 500 refereed papers in these areas.
Dr. Dobre serves as the VP Publications of the IEEE Communications Society. She was the inaugural Editor-in-Chief (EiC) of the IEEE Open Journal of the Communications Society and the EiC of the IEEE Communications Letters. 
Dr. Dobre was a Fulbright Scholar, Royal Society Scholar, and Distinguished Lecturer of the IEEE Communications Society. She obtained 7 IEEE Best Paper Awards including the 2024 Heinrich Hertz Award. Dr. Dobre is an elected member of the European Academy of Sciences and Arts, a Fellow of the Engineering Institute of Canada, and a Fellow of the Canadian Academy of Engineering. 

\end{IEEEbiography}

\end{document}